\documentstyle[prl,floats,aps]{revtex}

\begin{document}

\input{epsf.sty}

\draft

\twocolumn[\hsize\textwidth\columnwidth\hsize\csname @twocolumnfalse\endcsname

\title{Test-beds and applications for apparent horizon finders
in numerical relativity}

\author{M. Alcubierre${}^{1}$
\and S. Brandt${}^{1}$ 
\and B. Br\"{u}gmann${}^{1}$ 
\and C. Gundlach${}^{1}$ 
\and J. Mass\'{o}${}^{1,4}$ 
\and E. Seidel${}^{1,2,3}$ 
\and P. Walker${}^{1}$
}

\address{
${}^{(1)}$Max-Planck-Institut f\"ur Gravitationsphysik, 
Albert-Einstein-Institut, Schlaatzweg 1, 14473 Potsdam, Germany}

\address{
${}^{(2)}$ National Center for Supercomputing
Applications,
Beckman Institute, 405 N. Mathews Ave., Urbana, IL
61801
}
\address{
${}^{(3)}$ Departments of
Astronomy and Physics,
University of Illinois, Urbana, IL 61801
}

\address{
${}^{(4)}$ Departament de Fisica, Universitat de les Illes Balears, SPAIN
}
\date{\today}

\maketitle

\begin{abstract} We present a series of test beds for numerical 
  codes designed to find apparent horizons.  We consider three 
  apparent horizon finders that use different numerical methods: one 
  of them in axisymmetry, and two fully three-dimensional.  We 
  concentrate first on a toy model that has a simple horizon 
  structure, and then go on to study single and multiple black hole 
  data sets.  We use our finders to look for apparent horizons in 
  Brill wave initial data where we discover that some results 
  published previously are not correct.  For pure wave and multiple 
  black hole spacetimes, we apply our finders to survey parameter 
  space, mapping out properties of interesting data sets for future 
  evolutions.
\end{abstract}


\narrowtext

\vskip2pc]


\section{Introduction}

The ability to find black holes, if they exist, in numerically 
generated spacetimes is an important and difficult problem in 
numerical relativity.  Many algorithms have been developed over the 
years, but few have been tested extensively on numerically computed 
spacetimes, and even fewer have been applied to full three-dimensional 
(3D) spacetime constructions in Cartesian coordinates, which is the 
most common choice for 3D numerical relativity.  As we review below, 
black holes may exist through topological construction or high 
concentrations of mass-energy in the initial data, may form through 
collapse of matter or pure gravitational waves, and may move through 
the spacetime.  The dynamical and numerical properties of black holes 
in numerical relativity make this a very challenging problem, yet one 
that must be solved if we are to understand the physics of such 
simulations, or even if we are to evolve the systems for long periods 
of time.  In this paper we investigate and compare the properties of 
several algorithms developed to find black holes on an extensive 
series of analytic and numerically generated spacetimes.  We show that 
due to the sensitive nature of this problem, some previously published 
results on the existence of apparent horizons in numerically generated 
spacetimes are incorrect.  We also show examples where finders could 
pass certain test-beds, but failed on more complex cases, revealing 
coding errors, and also how in some cases existing algorithms had to 
be modified in order to locate difficult-to-find horizons.  Finally, 
we use these newly developed horizon finders to map out the parameter 
space of a series of black hole and gravitational wave data sets not 
yet studied in preparation for future numerical evolutions.

Black holes are defined by the existence of an event horizon (EH), the
surface of no return from which nothing, not even light, can escape.
The event horizon is the boundary that separates those null geodesics
that reach infinity from those that do not.  The global character of
such a definition implies that the position of an EH can only be found
if the whole history of the spacetime is known.  For numerical
simulations of black hole spacetimes in particular, this implies that
in order to locate an EH one needs to evolve sufficiently far into the
future, up to a time where the spacetime has basically settled down to
a stationary solution.  Recently, methods have been developed to
locate an analyze event horizons in numerically generated spacetimes,
with a number of interesting results
obtained~\cite{Anninos94f,Libson94a,Hughes94a,Matzner95a,Masso95a,Shapiro95a}.

In contrast, an apparent horizon (AH) is defined locally as the
outer-most marginally trapped surface~\cite{Hawking73a}, i.e. a
surface such that the expansion of out-going null geodesics is zero.
An AH can therefore be defined on a given spatial hypersurface.  A
well known result~\cite{Hawking73a} guarantees that if cosmic
censorship holds and an AH is found, then an EH must exist somewhere
outside of it and hence a black hole has formed.

Being able to find an AH has become a very important issue in the
numerical studies of black hole spacetimes.  An important reason for
this is the development of the so-called apparent horizon boundary
condition (AHBC), which excises singularity containing regions
interior to the AH in order to extend the evolution.  The basic idea
behind this is the fact that since the interior of a black hole is
causally disconnected from the rest of the spacetime, one can in
principle safely ignore it in a numerical evolution, thus avoiding the
problem of having to deal with the singularity through the use of a
pathological time slicing.  Several schemes are being currently
implemented to deal with such
AHBC~\cite{Seidel92a,Cook97a,Gundlach98a}.  The AH can also be used to
determine important information about the spacetime itself; its
topology (e.g., multiple disconnected 2--spheres) and its geometry
(e.g., its area and local curvature) provide important details of the
dynamics of the black holes present in the
spacetime~\cite{Anninos94f,Anninos93a,Anninos95c,Anninos93d}.  Here we
will not deal with such issues, but rather with the problem of finding
the AH in a reliable way in the first place.

Many different methods for locating AHs have been developed in the
past years, both for axisymmetric (2D) and fully 3D
spacetimes~\cite{Cadez74,Eppley77,Bishop82,Nakamura84,Cook90a,Kemball91a,Libson94b,Libson95a,Thornburg95,Baumgarte96,Gundlach97a,Shibata97a}.
Since even in exact black hole solutions the position of an AH
usually has to be determined numerically (with the exception of trivial
cases like a Schwarzschild or Kerr black hole), we have found it very
useful to develop several independent apparent horizon finders (AHF),
using different algorithms and even written by different people.
Comparing the results of these AHFs has allowed us both to make
careful studies of the horizon structures of a series of spacetimes,
and also to compare the performance of the different algorithms. Our
tests include simple toy spacetimes, single and multiple black hole
spacetimes, and pure gravitational wave spacetimes.  With respect to
the latter, we have in fact been able to show that some previously
published results are not correct.

When comparing the different AHFs we have concentrated both on the
reliability with which they can locate an AH and also on the speed
at which they can do this.  While speed is not such a fundamental
consideration when looking for AHs on one given spatial geometry
(it only needs to be done once, so one can afford to wait), it becomes
of crucial importance when trying to locate them on an evolving
spacetime, as will be required in any successful implementation of an
AHBC.  An AHF that takes much longer than an evolution step to locate
an AH can not be used very often without having a disastrous
impact on the performance of an evolution code.

A final word on our terminology:  Since in this paper we will not deal
with the problem of finding event horizons, from now on we will use
the term `horizon' to mean always a marginally trapped surface.  As we
will see in the examples below, there can often be more than one such
surface.  The AH will be by definition the outermost and we will refer
to those marginally trapped surfaces that lie inside it as `inner
horizons'.


\section{Finding Apparent Horizons}

\subsection{Basic Equations}

An AH is defined as the outer-most marginally trapped
surface~\cite{Hawking73a}, that is, a surface where the expansion of
out-going null geodesics vanishes.  In order to find a mathematical
expression for this definition, let us start by considering a smooth
spacelike hypersurface $\Sigma$ embedded in a spacetime
$(M,g_{\mu\nu})$ (in the following we will use the Greek alphabet to
denote spacetime indices on $M$, and the Latin alphabet to denote
spatial indices on $\Sigma$).  Let $\gamma_{ij}$ and $K_{ij}$ be the
induced 3-metric and extrinsic curvature of $\Sigma$, respectively.
Let now $S$ be a closed smooth two-dimensional surface embedded in
$\Sigma$, with unit outward pointing normal vector $s^{\mu}$.  The
expansion $H$ of a congruence of null rays moving in the outward
normal direction to $S$ can then be shown to be~\cite{York89}
\begin{equation}
H = \nabla_i s^i + K_{ij} s^i s^j - {\rm tr} K ,
\label{eqn:expansion}
\end{equation}
where $\nabla_i$ is the covariant derivative associated with the
3-metric $\gamma_{ij}$. An AH is then the outer-most surface such that
$H=0$ is everywhere on the surface (and the surface is smooth).

Let us now rewrite Eq.~(\ref{eqn:expansion}) by assuming that our
surface has been parameterized as a level set
\begin{equation}
F(x^i) = 0 .
\end{equation}

It is now straightforward to rewrite H in terms of the function $F$
and its derivatives.  First, we write the unit normal vector $s^i$ as
\begin{equation}
s^i = \frac{\nabla^i F}{|\nabla F|} .
\end{equation}

Substituting this in Eq.~(\ref{eqn:expansion}) gives us
\begin{equation}
H = \left( \gamma^{ij} - \frac{\nabla^i F \; \nabla^j F}{|\nabla F|^2} \right)
\left( \frac{\nabla_i \nabla_j F}{|\nabla F|} - K_{ij} \right) = 0 .
\label{eqn:horizon1}
\end{equation}
Equation~(\ref{eqn:horizon1}) is the basic equation to be solved when
looking for an AH.  That is, one must find the outermost closed
2-surface defined by $F(x^{i}) = 0$ such that Eq.~(\ref{eqn:horizon1})
is satisfied.


\subsection{Axisymmetric finder}

In axisymmetry, we find the surface by assuming that we have
a 2-sphere enclosing the origin.  We implement this by taking
\begin{equation}
F = r - R(\theta) ,
\end{equation}
and searching for the function $R(\theta)$ such that
Eq.~(\ref{eqn:expansion}) is zero when evaluated at $r=R(\theta)$.

Because our grid is symmetric across the axis, and because we know
that $F$ must be smooth across the axis in axisymmetry, we can easily
test whether an AH which intersects the axis at a particular value of
$z$ exists.  We simply substitute $F$ given above
into Eq.~(\ref{eqn:horizon1}) and integrate the resulting equation for $R$
from the axis using $R(\theta=0)=z$ and $\partial_\theta
R(\theta=0)=0$ as boundary conditions.  When we reach $\theta_{max}$
we calculate how closely we have come to satisfying the condition
$\partial_\theta R(\theta=\theta_{max})=0$ ($\theta_{max}$ is either
$\pi/2$ or $\pi$ depending on whether we have chosen equatorial plane
symmetry or not).  We integrate $R$ using many different starting
values (i.e. values of $z$), and search for two neighboring values
which bracket the condition at $\theta_{max}$.  Finally, we bisect
until we reach the desired precision.

This reduces the process of searching for the AH to a one parameter
search, namely the search for the proper $z$ value at which to start
the integration.  Because the search space has been so greatly
reduced, we can have a high degree of confidence that we have found
{\em the} AH and not merely a trapped surface.


\subsection{3D minimization algorithm}

Minimization algorithms for finding AH were among the first methods
ever tried.  They were in fact the original methods used by Brill and
Lindquist~\cite{Brill63} and by Eppley~\cite{Eppley77}.  More recently,
a 3D minimization algorithm was developed and implemented by the
NCSA/WashU group, applied to a variety of black hole initial data and
3D numerically evolved black hole
spacetimes~\cite{Libson94b,Libson95a,Libson93a,Camarda97a,Camarda97c}.
Essentially the same algorithm was also implemented independently by
Baumgarte {\em et al.}~\cite{Baumgarte96}.

The basic idea behind a minimization algorithm is to expand the
parameterization function $F(x^i)$ in terms of some set of basis
functions, and then minimize the integral of the square of the
expansion $H^2$ over the surface.  At an AH this integral should
vanish, and we will have a global minimum.  Of course, since
numerically we will never find a surface for which the integral
vanishes exactly, one must set a given tolerance level below which a
horizon is assumed to have been found.  The only way to be certain
that this is a true horizon is to check if the value of the integral
keeps diminishing when we increase either the resolution of the
numerical grid, or the number of terms in the spectral decomposition.

Minimization algorithms for finding AHs have a few drawbacks: First,
the algorithm can easily settle down on a local minimum for which the
expansion is not zero, so a good initial guess is often required.
Moreover, when more than one marginally trapped surface is present (as
will be the case in several of the spacetimes considered here) it is
very difficult to predict which of these surfaces will be found by the
algorithm: The algorithm can often settle on an inner horizon instead
of the true AH.  Again, a good initial guess can help point the finder
towards the AH.  Notice that for time-symmetric data one can usually
overcome this problem by looking for a minimum of the area instead,
since for vanishing extrinsic curvature the AH will be an extremal,
and generically a minimal, surface (of course, one can always think of
cases where there is more than one minimal surface, so we would still
need a good initial guess).  Finally, minimization algorithms tend to
be very slow when compared with `flow' algorithms of the type
described in the next section.  Typically, if $N$ is the total number
of terms in the spectral decomposition, a minimization algorithm
requires of the order of a few times $N^2$ evaluations of the surface
integrals (where in our experience `a few' can sometimes be as high as
10).  Since the number of terms in the decomposition is
$(l_{max}+1)^2$, the total time required grows as ${l_{max}}^4$ (so
eliminating as many terms as possible making use of whatever
symmetries there might be in our data can have an enormous impact on
the speed of the algorithm).

For the specific minimization algorithm that we have used for this
work we start by parameterizing the surface in the following way:
\begin{equation}
F(r,\theta,\phi) = r - h(\theta,\phi) .
\label{eqn:F}
\end{equation}
The surface under consideration will be taken to correspond to the
zero level of $F$.  The function $h(\theta,\phi)$ is then expanded
in terms of spherical harmonics:
\begin{equation}
h(\theta,\phi) = \sum_{l=0}^{l_{\rm max}} \sum_{m=-l}^{l}
\sqrt{4\pi} \, a_{lm} Y_{lm}(\theta,\phi) .
\label{eqn:harmonics}
\end{equation}
The overall factor of $\sqrt{4\pi}$ has been inserted so that $a_{00}$
is the average (coordinate) radius of the surface, $a_{10}$ is its
average displacement in the $z$-direction, and so on. We also use a
real basis of spherical harmonics, such that $m$ and $-m$ stand for an
angular dependence $\cos(m\phi)$ and $\sin(m\phi)$ instead of
$\exp(im\phi)$ and $\exp(-im\phi)$.

Given a trial function $h$, we construct $F$ using Eq.~(\ref{eqn:F})
and calculate the expansion $H$ as a 3D function from
Eq.~(\ref{eqn:horizon1}) using finite differences.  We interpolate $H$
onto a two-dimensional grid in $\{\theta,\phi\}$ at those points where
$r = h(\theta,\phi)$.  Finally, we calculate the surface integral of
$H^2$.  We then use a standard minimization algorithm (Powell's method
in multi-dimensions~\cite{Press86}) to find the values of the
coefficients $a_{lm}$ for which the integral reaches its minimum.

Once we find a candidate horizon (that is, a surface for which the
integral of $H^2$ is below a certain threshold), we typically increase
the number of terms in the spherical harmonics expansion, and the 3D
spatial resolution of our numerical grid to see if the integral keeps
diminishing.  If it does, the surface is a real horizon, but if it
reaches a limiting value different from zero, it is just a local
minimum of $H^2$.  In our experience this procedure usually works very
well except when we are in a situation where we are close to a critical
value of some parameter for which an AH first forms.  In such a case,
the exact critical value can not be determined very accurately because
of the inability of the algorithm to distinguish between a real
horizon and a very low local minimum.

This algorithm has been implemented in the Cactus code for 3D
numerical relativity~\cite{Bona98b}, which is used to compute the 3D
results for the present paper.  For more details on the application of
this algorithm see
Refs.~\cite{Libson94b,Libson95a,Baumgarte96,Libson93a}.


\subsection{3D fast flow algorithm}

A second method that has been implemented in the Cactus code is the
``fast flow'' method proposed by Gundlach~\cite{Gundlach97a}.  The
fast flow algorithm describes the horizon using the same
function~(\ref{eqn:F}) and the same decomposition in spherical
harmonics~(\ref{eqn:harmonics}) as described above. Here we do not
discuss why this method is expected to work, but limit ourselves to 
a brief definition of the algorithm, followed by a few comments.

Starting from an initial guess for the $a_{lm}$, typically one
representing a large sphere inscribed into the numerical domain, the
algorithm approaches the AH through the iteration procedure defined by
\begin{equation}
\label{SpectralAHF}
a_{lm}^{(n+1)} 
= a_{lm}^{(n)}
-{A\over 1 + B l(l+1)}
\left(\rho H\right)_{lm}^{(n)}.
\end{equation}
where $(n)$ labels the iteration step, $\rho$ is some positive
definite function (``a weight''), and $(\rho H)_{lm}$ are the Fourier
components of the function $\rho H$. Various choices for the weight
function $\rho$ and the constant coefficients $A$ and $B$ parameterize
a family of such methods. Here we use the specific weight
\begin{equation}
\label{rho}
\rho = 2 \; r^2 |\nabla  F| \left[ \left(g^{ij}-s^i s^j \right)
\left( \bar g_{ij}-\nabla_i r \nabla_j r \right) \right]^{-1} ,
\end{equation}
where $\bar g_{ij}$ is the flat background metric associated with the
coordinates $(r,\theta,\phi)$, or $(x,y,z)$. 
We use values of $A$ and $B$ that depend on $l_{\rm max}$ through
\begin{equation}
\label{alphabeta}
A = {\alpha\over l_{\text{max}}(l_{\text{max}}+1)} + \beta,
\quad
B = {\beta\over \alpha}.
\end{equation}
Here, we have used $\alpha=c$ and $\beta=c/2$, where $c$ is a variable
step size, with a typical value of $c\sim 1$. The iteration procedure
can clearly be seen as a finite difference approximation to a
parabolic flow, and the adaptive step size is chosen to keep the
finite difference approximation roughly close to the flow limit to
prevent overshooting of the true apparent horizon. The adaptive step
size is determined by a standard method used in ODE integrators: we
take one full step and two half steps and compare the resulting
$a_{lm}$. If the two results differ too much one from another, the
step size is reduced.

The motivation for and history of this ansatz is discussed
in~\cite{Gundlach97a}. Here we limit ourselves to a few isolated
comments to indicate how the method relates to other methods. The
method is clearly related to Jacobi's method for solving an elliptic
equation, here the elliptic equation $H=0$, by transforming it into a
related parabolic equation. Going from the discrete equation
(\ref{SpectralAHF}) to the continuum limit defined by $c\to 0$, $nc\to
\lambda$ and $l_{\text{max}}\to\infty$, with the continuous parameter
$\lambda$ replacing the discrete iteration number $n$, we obtain
\begin{equation}
{\partial h(\theta,\phi,\lambda) \over \partial \lambda} =
- \left( 1 + {\beta\over \alpha} L^2\right)^{-1} \rho[h] \, H[h].
\end{equation}
The method is called a ``fast'' flow for $\beta > 0$ because the
division by $l(l+1)$, corresponding to the inverse of the Laplace
operator $L^2$ on the sphere, allows us to take very large steps in
$\lambda$ even using an explicit difference method. Very large here
means that the number of iteration steps to convergence is between 10
and 100, many fewer than the number of grid points on the horizon.  A
method that would be called ``slow flow'' in our terminology, defined
by $\beta = 0$ and $\rho = |\nabla F|$, was proposed by Tod
\cite{Tod91}. It reduces to curvature flow for $K_{ab}=0$.  It should
finally be mentioned that the particular choice of $\rho$ used here is
motivated by the AHF algorithm of Nakamura et al.~\cite{Nakamura84},
but that $\rho=1$ and $\rho = |\nabla F|$ are workable choices, too.



\section{Toy ``bowl'' spacetimes}

As a first test of our AHFs we consider a set of ``toy'' spacetimes
which do not satisfy the Einstein equations, but which contain
horizons.  We chose spacetimes that have the classical ``bag of gold''
geometry, {\em i.e.} they have a maximal and a minimal surface with an
essentially flat interior, and an asymptotically flat exterior.  We
will refer to these spacetimes as the ``bowl spacetimes''.  Here we
will only consider static versions of these spacetimes but we should
mention that time dependent generalizations where the geometry starts
from flat space and ``collapses'' smoothly to a bag of gold are easy
to construct.

Our ansatz for the spacetime metric is
\begin{equation}
ds^2 = - dt^2 + dr^2 + \left[ r - \lambda f(r) \right]^2
d\Omega^2 , \quad 0 \leq \lambda f < r,
\label{eqn:gaussbowl}
\end{equation}
where $\lambda$ is the bowl strength parameter, and $d\Omega$ is the 
standard flat space differential solid angle.  We consider two classes 
of static bowl metrics, the Gaussian bowl,
\begin{equation}
f(r) = f_{\rm G} = e^{-(r-r_0)^2/\sigma^2},
\end{equation}
and the Fermi bowl,
\begin{equation}
f(r) = f_{\rm F} = \frac{1}{1 + e^{-\sigma (r-r_0)}}.
\end{equation}
Notice that the bowl spacetimes defined above are not completely 
regular at the origin.  This could of course be fixed by choosing 
instead of the Gaussian and Fermi functions some other functions that 
satisfied the appropriate boundary conditions at the origin.  We feel, 
however, that this is not really necessary as long as we use values of 
$\sigma$ and $r_0$ such that $f_{\rm G}$ and $f_{\rm F}$ have very 
small values at $r=0$.

Each of these metrics has an extremal surface when
the following condition is satisfied
\begin{equation}
\frac{dg_{\theta\theta}}{dr} = 0 .
\label{eqn:horizon2}
\end{equation}
Since we have imposed $K_{ij}=0$, it is easy to see that this is
equivalent to the condition for the existence of a
horizon Eq.~(\ref{eqn:horizon1}).

\begin{figure}
\epsfysize=3in \epsfbox{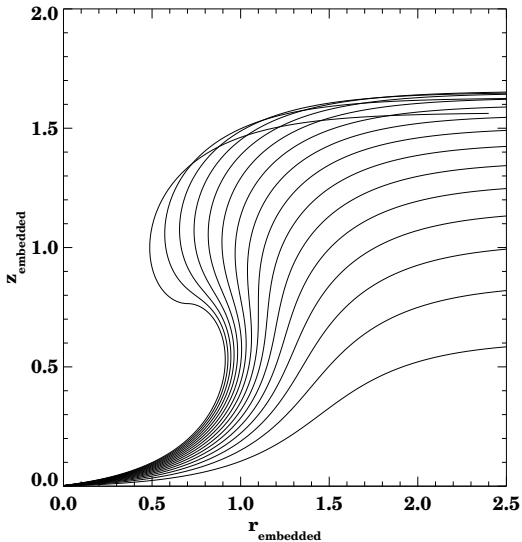}
\caption{We show Fermi bowl embeddings for values of $\lambda$ between
  $0$ and $1.6$ every $0.1$, with $r_0 = 1.5$ and $\sigma = 5.0$. The
  ``bag of gold'' feature of the higher $\lambda$ metrics is clear
  from the embedding.}
\label{fig:fermibowl}
\end{figure}

The Fermi bowl has an advantage over the Gaussian bowl in that it is
embeddable in a fictitious Euclidean flat space for a wide range of
$\lambda$, $\sigma$ and $r_0$, while the Gaussian bowl is not.  The
reason for this is that for large $r$, the angular metric of the
Gaussian bowl approaches $r^2$ too quickly.  Since $r$ always measures
proper radial distance, we find ourselves in a situation where, far
away, areal radius coincides with distance to the origin.  As we come
in from infinity, the deviations in $g_{\theta\theta}$ from its flat
values try to force the embedding away from a flat slice, but this can
not happen because our distance to the origin would then be
significantly affected, so the geometry is not embeddable. The Fermi
bowl, on the other hand, avoids this problem by instead having an
asymptotic angular metric of the form $(r-\lambda)^2$, which means
that far away the distance to the origin is larger than the areal
radius.  This ``extra'' distance gives the embedding the room it needs
to accommodate the changes in $g_{\theta \theta}$ in the region $r
\simeq r_0$.  In Fig.~\ref{fig:fermibowl} we show the embeddings of
the Fermi bowl metric for the parameters $r_0 = 1.5$ and $\sigma =
5.0$, for a range of $\lambda$ between $0$ and $1.6$. From this figure
it is intuitively clear why larger $\lambda$ bowl spacetimes have
minimal surfaces at the ``neck'' of the bag of gold.

For our purposes here, we rewrite the bowl metrics in Cartesian
coordinates as
\begin{eqnarray}
ds^2 &=&  \left[ x^2 + \left( 1 - \lambda f / r
\right)^2 \left( y^2 + z^2 \right) \right] \frac{dx^2}{r^2} \nonumber \\
&+&  \left[ y^2 + \left( 1 - \lambda f / r
\right)^2 \left( x^2 + z^2 \right) \right] \frac{dy^2}{r^2} \nonumber \\
&+&  \left[ z^2 + \left( 1 - \lambda f / r
\right)^2 \left( x^2 + y^2 \right) \right] \frac{dz^2}{r^2} \nonumber \\
&+& \frac{\lambda f}{r^3} \left( 2 - \lambda f / r \right) 
\left( x y dx dy + x z dx dz + y z dy dz \right) .
\label{eqn:bowlcartesian}
\end{eqnarray}
Notice that even though the bowl spacetimes
are spherically symmetric by construction, we can hide this symmetry
from the AHF by a simple rescaling of the coordinates
\begin{equation}
x' = d_x x , \quad y' = d_y y , \quad z' = d_z z .
\end{equation}
(The $d_{i}$ are scaling constants, not differential operators.)
Let us now return to the condition for the existence of a horizon
Eq.~(\ref{eqn:horizon2}).  From the form of the metric it is clear
that this condition takes the simple form
\begin{equation}
\lambda f'(r) = 1 .
\label{eqn:horizon3}
\end{equation}
For the Fermi bowl, this equation can in fact be solved in closed form
to find
\begin{equation}
r = r_0 - \frac{1}{\sigma} \; \ln \left\{ \left( \lambda \sigma / 2
- 1 \right) \pm \left( \lambda^2 \sigma^2 / 4 - \lambda \sigma
\right)^{1/2} \right\} .
\end{equation}
Notice that from this we can easily see that a horizon will first
appear for
\begin{equation}
\lambda = 4 / \sigma , \qquad r = r_0 .
\end{equation}

For the Gaussian bowl condition~(\ref{eqn:horizon3}) can not be solved
in closed form but it can be solved numerically to arbitrary high
accuracy by simple bisection.  Doing this we find that if we take for
example $r_0 = 2.5$ and $\sigma = 1$, for the Gaussian bowl an AH
first appears for \mbox{$\lambda = 1.165$}, while for the Fermi bowl
it appears for \mbox{$\lambda = 4$}.  For larger values of $\lambda$,
we can also tabulate the positions of the horizons.
Table~\ref{tab:bowl} shows the coordinate radius of the AH and the
inner horizon for different values of $\lambda$ for the Gaussian and
Fermi bowls.

\begin{table}
\begin{tabular}{c|c|c}
\multicolumn{3}{c}{Gaussian bowl} \\ \hline
$\lambda$ & $r$ inner horizon & $r$ apparent horizon \\ \hline
1.165 & no   & 1.79 \\
1.250 & 1.60 & 1.97 \\
1.500 & 1.41 & 2.11 \\
1.750 & 1.30 & 2.18 \\
2.000 & 1.22 & 2.23 \\
2.250 & 1.16 & 2.26 \\
2.393 & 1.13 & 2.28 (pinch-off) \\ \hline \hline
\multicolumn{3}{c}{Fermi bowl} \\ \hline
$\lambda$ & $r$ inner horizon & $r$ apparent horizon \\ \hline
4.000 & no   & 2.50 \\
4.250 & 2.00 & 2.99 \\
4.500 & 1.81 & 3.19 \\
4.750 & 1.66 & 3.34 \\
4.782 & 1.64 & 3.35 (pinch-off) 
\end{tabular}
\caption{Coordinate radius of the inner and apparent horizons for different
values of $\lambda$ for the Gaussian and Fermi bowl spacetimes in the
particular case when $r_0 = 2.5$ and $\sigma = 1$. In each case, we
consider values of $\lambda$ between the first appearance of a horizon
and the pinch-off, {\em i.e.} the value of $\lambda$ for which
$g_{\theta\theta}$ first has a zero.}
\label{tab:bowl}
\end{table}

Our first test was to see how well our AHFs could reproduce the 
results found by solving Eq.~(\ref{eqn:horizon3}) in the spherically 
symmetric case.  This might seem like a very trivial test, but it was 
while studying this case that we discovered a weakness of the original 
implementation of the fast flow algorithm.  This weakness was not a 
programming error, but rather an unexpected consequence of the speed 
at which this algorithm can proceed.  We discovered that for 
spherically symmetric bowl spacetimes that had a narrow (but still 
very evident) region of trapped surfaces, the algorithm would just 
jump over the whole trapped region and conclude that there was no AH. 
The reason for this was that the step-size used by the algorithm was 
just too big.  Our first attempt at a cure was to reduce the step-size 
by hand, but this made the finder very slow and defeated the whole 
idea of a ``fast flow''.  A much better solution in the end was to 
implement an adaptive step-size routine as part of the algorithm.  We 
should stress the point that this type of situation, where we have a 
narrow shell of trapped surfaces with essentially flat space both 
inside and out, can in fact occur in real physical systems such as the 
Brill wave spacetimes that will be discussed below.  Although the 
constant step size code had passed various other test-beds involving 
black holes, this simple case led to important modifications of the 
algorithm that make the crucial difference between success and 
failure of the method.

Apart from the problem we just mentioned, all three AHFs performed
very well in the spherically symmetric case, finding the horizons in
the correct positions, and finding also the correct critical values of
$\lambda$ for which an AH first forms.  For example, for the Gaussian
bowl with \mbox{$r_0=2.5$} and \mbox{$\sigma=1$}, the 2D finder
determined the value of $\lambda$ for which a horizon first appears to
be $\lambda_*=1.166$ (working on a $200\times100$ $\{r,\theta\}$
grid), while the 3D finders determined it to be in the interval
$\lambda_*\in[1.16,1.17]$.

In order to have a more interesting test, we will now rescale the $z$
axis to have what in coordinate space will appear to be an
axisymmetric spacetime.  This will still allow us to compare all three
AHFs. Later we will rescale both the $y$ and $z$ axis (with different
scaling factors) to compare the minimization and fast flow algorithm
in a fully 3D situation.  For the axisymmetric case we have considered
both oblate and prolate configurations.  In all cases we have studied
the AHFs still find the correct critical values of $\lambda$ with
about the same precision as in the spherical case.

\begin{figure}
\epsfysize=3in \epsfbox{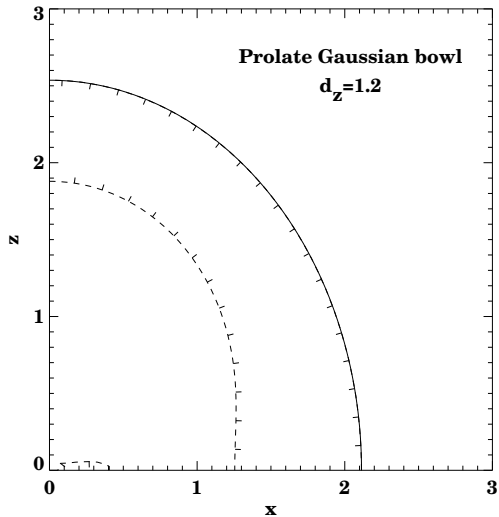}
\caption{Position of the AH on the $x-z$ plane
  for a prolate Gaussian bowl with $\lambda=1.5$, $r_0=2.5$,
  $\sigma=1$ and $d_z=1.2$. The solid line is the AH, and the dotted
  lines are the zeroes of the expansion $H$.}
\label{fig:gauss1}
\end{figure}

In Fig.~\ref{fig:gauss1} we show a visual test of the accuracy of the
position of the AH (found in this case with the minimization algorithm
with $l_{max}=8$) on the $x-z$ plane for a prolate Gaussian bowl with
$\lambda=1.5$, $r_0=2.5$, $\sigma=1$ and $d_z=1.2$.  To check if the
candidate surface is really a horizon, we can evaluate the residual
value of the expansion $H$ on the numerically found horizon, but we
also found the following more detailed test helpful in distinguishing
spurious from real apparent horizons.  Consider {\it all} the level
sets of the function $F(x)$.  The level set $F=0$ corresponds to our
candidate horizon, and must have $H=0$ everywhere up to numerical
error (i.e., the actual horizon surface must have {\it both} $F(x)=0$
{\it and} $H=0$).  On each of the other level sets, we should
generically still be able to find lines on which $H=0$, separating
regions with $H>0$ and $H<0$.  Linking up these lines on different
level sets, we obtain a set of surfaces that we call ``zeroes of the
expansion''.  Note that these surfaces have no geometric meaning, but
depend both on the coordinate system and the candidate AH.  A true AH
must coincide with one of these surfaces (numerically it should follow
one closely), while spurious AHs tend to be intersected by them.  In
our 2D plots, the solid line corresponds to the position of the AH
$F(x^i)=0$, while the dotted lines correspond to the zeroes of the
expansion $H$ on the level surfaces of $F$, as just described.  The
tick marks point in the direction of decreasing expansion, that is,
towards the trapped regions.  For this particular run we used a grid
with $80^3$ points and a resolution of \mbox{$\Delta x = \Delta y =
  \Delta z = 0.05$}.

To quantify more the agreement between our two 3D finders, in
Table~\ref{tab:gauss1} we compare the spectral coefficients found
which each finder for two different resolutions.  From the table we
can see how the coefficients found with the different algorithms are
closer for the higher resolution.  The only exception is the
coefficient of the $l=8$ term for which the difference almost doubled
(while still being only off about 3\%).  A small difference is not
surprising, though, since this is the last coefficient in the
spectral decomposition, and there is no reason why the errors
associated with the truncation of the infinite series should be the
same for both algorithms (they have very different termination
criteria).  In fact, if we increase $l_{max}$ to 12 we find that the
agreement in the $l=8$ coefficient improves dramatically.

\begin{table}
\begin{tabular}{c||c|c||c|c}
$l$
&
\multicolumn{2}{c||}{$40^3$, $\Delta x = 0.10$} 
&
\multicolumn{2}{c}{$80^3$, $\Delta x = 0.05$} 
\\ \hline
 & fast flow & minim.\ & fast flow & minim.\   
\\ \hline
0 & 2.24 & 2.24 & 2.24 & 2.24 \\
2 & $1.20 \times 10^{-1}$ & $1.20 \times 10^{-1}$ & $1.20 \times 10^{-1}$ &
$1.20 \times 10^{-1}$ \\
4 & $8.37 \times 10^{-3}$ & $8.40 \times 10^{-3}$ & $8.37 \times 10^{-3}$ &
$8.38 \times 10^{-3}$ \\
6 & $6.40 \times 10^{-4}$ & $6.50 \times 10^{-4}$ & $6.40 \times 10^{-4}$ &
$6.41 \times 10^{-4}$ \\
8 & $5.07 \times 10^{-5}$ & $4.97 \times 10^{-5}$ & $5.12 \times 10^{-5}$ &
$4.94 \times 10^{-5}$
\end{tabular}
\caption{Spectral coefficients defining the apparent horizon for
  a prolate Gaussian bowl with $\lambda=1.5$, $r_0=2.5$, $\sigma=1$
  and $d_z=1.2$, using $l_{max} = 8$.}
\label{tab:gauss1}
\end{table}

\begin{figure}
\epsfysize=3in \epsfbox{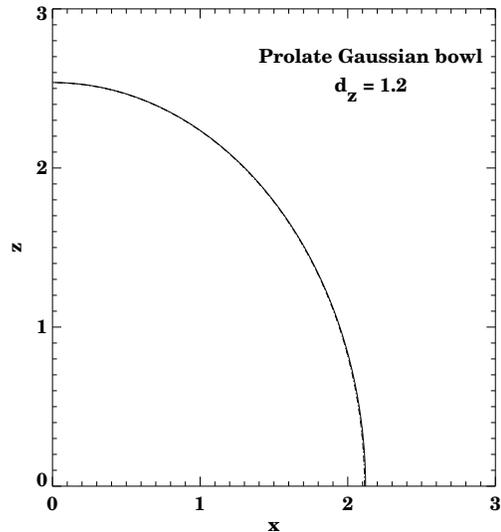}
\caption{Comparison of the AHs found with our three finders
  for a prolate Gaussian bowl with $\lambda=1.5$, $r_0=2.5$,
  $\sigma=1$ and $d_z=1.2$.  The solid line corresponds to the 2D
  finder, the dotted line to the fast flow finder, and the dashed line
  to the minimization finder.  All three lines lie on top of each
  other.}
\label{fig:gauss2}
\end{figure}

Fig.~\ref{fig:gauss2} shows a comparison of the horizons found with
our different finders for the same Gaussian bowl as above. Notice how
all three finders have found the same surface. We have also studied
many more axisymmetric configurations, with both prolate and oblate
horizons, and had found similar results.

Up until now we have only discussed the reliability of the different 
AHFs in locating the horizons.  As was mentioned in the introduction, 
another important question is how fast can horizons be located.  For 
the particular example considered above the 2D finder is very fast.  
For example, for a solution of the initial data on an $800^2$ 
$(\rho,z)$ grid, and an initial search of $100$ different starting 
points along the $z-$axis (which are then bisected to a precision of 
$10^{-8}$ for the exact $z$ value) the code took about 1 minute on a 
single processor on an SGI Origin 2000 parallel computer.  The 3D 
finders, not surprisingly, are much slower.  The minimization 
algorithm was in fact somewhat faster than the fast flow method, 
taking $\sim400$ seconds against the $\sim600$ seconds of the fast 
flow algorithm when running on 8 processors on the same machine as 
above.  This is somewhat deceptive, however, since the minimization 
algorithm was running in axisymmetric mode, that is, even though it 
worked on a full 3D grid it did not consider any $m\neq0$ terms in the 
expansion, while the fast flow algorithm considered all the terms.

As a final test, we have considered several non-axisymmetric
configurations.  Here we report results for the particular case of a
Fermi bowl with $\lambda=4.25$, $r_0=2.5$, $\sigma=1$, $d_y=1.2$ and
$d_z=0.8$.  For this run we used a grid with $80^3$ points and a
resolution of \mbox{$\Delta x = \Delta y = \Delta z = 0.06$}.  This
example illustrates clearly the two main drawbacks of the minimization
algorithm: in the first place it is considerably slower than the fast
flow algorithm, taking $\sim2000$ seconds against the $\sim300$
seconds of the fast flow (running again in the same machine and with
the same configuration as above), and in the second place the
minimization algorithm had a strong tendency to lock onto the inner
horizon instead of the AH. Since the bowl metrics are static, we could
overcome this problem by minimizing the area instead of the expansion,
which allowed us to lock onto the real AH (in general one could
presumably also have more that one minimal surface but this is not the
case in this example).  Notice that if after minimizing the area one
gives the final $a_{lm}$ coefficients as initial guess and tries to
minimize the expansion $H$ instead, the algorithm will not wander
anymore to the inner horizon but will instead stay in the AH, so the
problem is just one of having a good initial guess.  In an actual
evolution the horizon location of the previous find can be used as an
initial guess~\cite{Anninos94c}, but if the horizon spontaneously
jumps out or changes topology, as can happen when black holes are
highly distorted or merge, this will be of little value.

One might wonder why the fast flow algorithm is faster in this case
than in the axisymmetric configuration studied above.  The reason is
that for this configuration, the AH is closer to the edge of the
computational domain (and therefore closer to the initial trial
sphere) and the finder converges to it sooner.

\begin{figure}
\epsfysize=3.4in \epsfbox{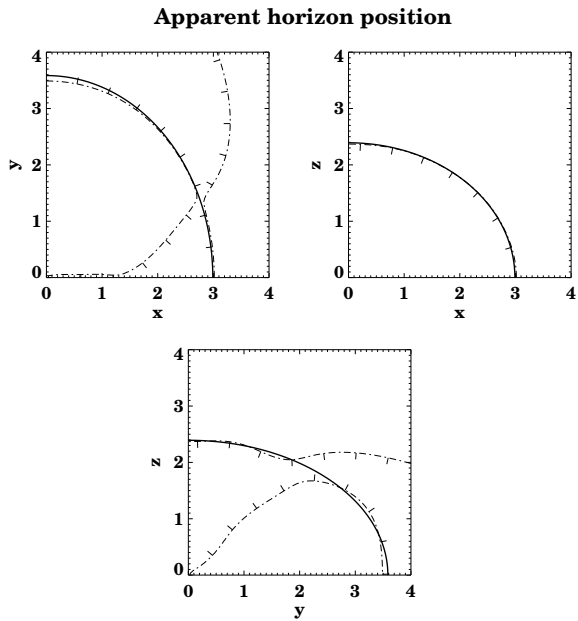}
\caption{Position of the AH on the $x-y$, $x-z$ and $y-z$ planes
  for a Fermi bowl with $\lambda=4.25$, $r_0=2.5$, $\sigma=1$,
  $d_y=1.2$ and $d_z=0.8$. The solid line is the AH, and the dotted
  lines are the zeroes of the expansion $H$.}
\label{fig:fermi1}
\end{figure}

\begin{figure}
\epsfysize=3.4in \epsfbox{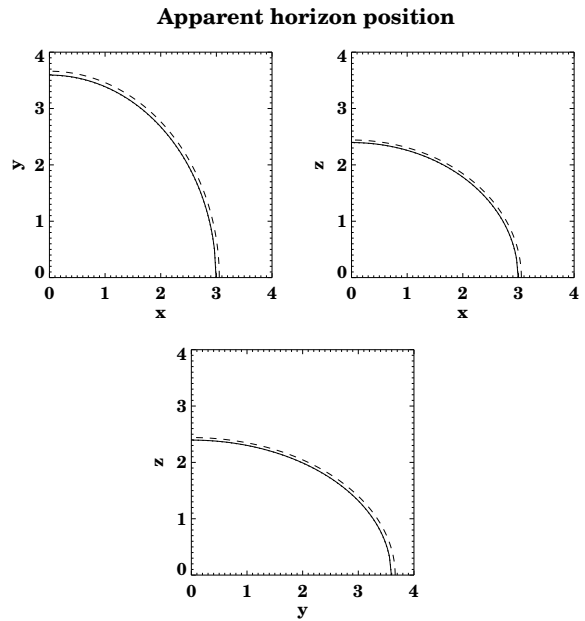}
\caption{Comparison of the AHs found with our two 3D finders for
  a Fermi bowl with $\lambda=4.25$, $r_0=2.5$, $\sigma=1$, $d_y=1.2$
  and $d_z=0.8$, using $l_{max}=8$. The solid lines correspond to the
  exact position of the horizon, the dashed lines to the position
  found with the fast flow finder, and the dotted lines to that found
  with the minimization finder.}
\label{fig:fermi2}
\end{figure}

Fig.~\ref{fig:fermi1} shows again a visual test of the accuracy of
the position of the AH, as found with the minimization algorithm with
$l_{max}=8$, on the $x-y$, $x-z$ and $y-z$ planes.  Again, the solid
lines correspond to the position of the horizon and the dotted lines
to the zeroes of the expansion $H$. The fact that the solid lines
coincide with a dotted line indicates that we have a true marginally
trapped surface.  In Fig.~\ref{fig:fermi2} we compare the two
different 3D AHFs.  The solid lines correspond to the exact position
of the horizon, the dashed lines to the position found using the fast
flow finder, and the dotted lines to that found with the minimization
finder.  Again, we have a very good agreement, though not as
impressive as the one found in the axisymmetric case. The minimization
finder has found the correct horizon to high accuracy, but the fast
flow finds a surface somewhat outside. This seems to be a general
property of our implementation of the fast flow algorithm: it has a
tendency to stop slightly outside the real horizon if $l_{max}$ is
not large enough.  If we use $l_{max}=12$ instead, keeping the same
spatial resolution, the three lines become indistinguishable.

\section{Black hole data}

We now turn to Cauchy data that contain a black hole by construction.
These initial data have throats that either connect two asymptotically
flat sheets identified by an isometry operator (Misner type data), or
that connect one asymptotically flat region to as many asymptotically
flat sheets as there are black holes (Brill-Lindquist type data).
More important than the difference in the topology of the initial data
slice is whether the initial data is time symmetric or not, and we
discuss both cases below.  Some of these data sets are known
analytically, while others can be computed only by solving the
Hamiltonian and momentum constraints.  In either case, the question is
not whether a black hole exists, but rather where is the apparent
horizon, and how many components does it have.  We apply the apparent
horizon finders described above to several of these spacetimes and
compare their ability to find the correct AH surfaces.

\subsection{Time symmetric black hole data}


\subsubsection{Two black hole data}

There are a number of two black hole initial data sets of interest,
and here we will consider the two must commonly used in numerical
relativity known respectively as the Misner and the Brill-Lindquist
data sets.

The classic two black hole spacetime considered over several
generations of numerical relativists is provided by the Misner data
for time-symmetric, axisymmetric, equal mass black
holes~\cite{Cadez74,Misner60,Hahn64,Smarr76,Anninos93b,Anninos94b}.
The black holes in the Misner data are connected via throats to a
single asymptotically flat universe.  The horizon structure of this
initial data system has been well studied, and provides important
tests of a horizon finder's ability to distinguish between spacetimes
with with a single distorted black hole horizon or one with two
disjoint apparent horizons.

The Misner initial data are parameterized by a distance parameter 
$\mu$, related to the proper distance between two throats.  

Misner's 3-metric takes the conformally-flat form~\cite{Misner60}
\begin{equation}
d\ell^2 = \psi^4 \left( dx^2+dy^2+dz^2 \right),
\label{confla}
\end{equation}
with the conformal factor $\psi$ given by
\begin{equation}
\psi = 1 + \sum^{\infty}_{n=1} \frac{1}{\sinh \left( n\mu \right)}
\left( \frac{1}{{}^+r_n} + \frac{1}{{}^-r_n} \right),
\end{equation}
where
\begin{equation}
\label{eq:rn}
{}^{\pm}r_n = \sqrt{x^2 + y^2 + \left( z \pm \coth \left( n\mu \right)
    \right)^2}.
\end{equation}

\begin{figure}
\epsfysize=3in \epsfbox{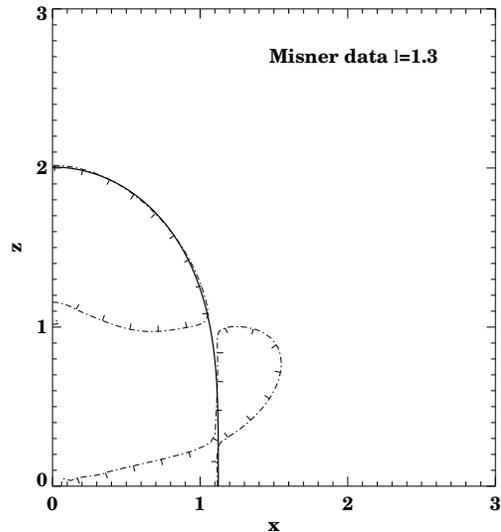}
\caption{Position of the AH on the $x-z$ plane
  for Misner data with $\mu=1.4$. The solid line is the AH, and the
  dotted lines are the zeroes of the expansion $H$.}
\label{fig:misner1}
\end{figure}

The black holes are on the $z$-axis, with their centers at
$z=\pm\coth\mu$ and with throat radii $a=1/\sinh\mu$.  As $\mu$ is
increased, the centers of the holes approach each other in coordinate
space, and their throat radii decrease.  The net effect is that larger
values of $\mu$ correspond to the throats being farther away from each
other in proper distance, scaled by the ADM mass $M$.  Studies have
shown that beyond a certain critical value $\mu_*=1.36$ the system
goes from a single horizon to two disjoint horizons~\cite{Cadez74}.
Both the 3D finders, working with a grid spacing of $\Delta x = \Delta
y = \Delta z = 0.06$ and $l_{\rm max}=4$, find a common AH in the
Misner data at $\mu=1.3$, but not at $\mu=1.4$ even after doubling
$l_{\rm max}$ and halving $\Delta x$. The axisymmetric finder, using a
grid spacing of $\Delta r=0.00375$ and 400 angular zones, finds a much
more precise critical value of $\mu_*=1.364$.  Fig.~\ref{fig:misner1}
shows our standard visual test for the position of the horizon for the
case $\mu=1.3$ (as found with the minimization algorithm with $l_{\rm
  max}=8)$.  Notice how the horizon does coincide with a zero of the
expansion. We have also calculated the horizon area as a
coordinate-independent quantity.  For the case $\mu=1.3$ we find an
area of $A \sim 5.0\times 10^2$.

\begin{figure}
\epsfysize=3in \epsfbox{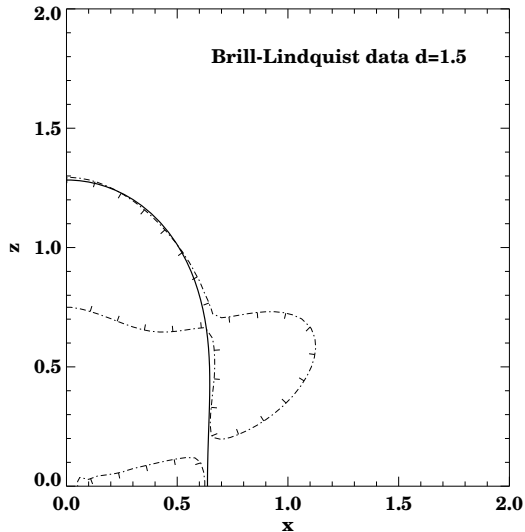}
\caption{Position of the AH on the $x-z$ plane
  for Brill-Lindquist data with $d=1.5$. The solid line is the AH,
  and the dotted lines are the zeroes of the expansion $H$.}
\label{fig:brilin1}
\end{figure}

The Brill-Lindquist~\cite{Brill63} initial data describe a
time-symmetric slice in which each black hole is connected via a
throat to its own asymptotically flat region. In contrast, both black
holes in Misner data are connected to the {\it same} opposite
asymptotically flat region, so that the slice is multiply
connected. Misner data therefore have an additional reflection-type
isometry that is absent in Brill-Lindquist data. Mathematically,
Brill-Lindquist data can be described as taking only the first term in
the infinite sum leading to Misner data. Brill-Lindquist data for two
black holes form a one-parameter family, but instead of by $\mu$ they
are more commonly parameterized by keeping the naked masses of the two
black holes fixed (the naked masses are the ADM masses of the
disconnected asymptotic regions) and varying their coordinate distance
$d$. In terms of this coordinate distance, the critical distance for
two black holes of naked mass one each is given as $d_*=1.56$ by Brill
and Lindquist~\cite{Brill63}, and $d_*=1.53$ by Bishop {\em et
al.}~\cite{Bishop82}.  Both our 3D finders, working with a grid
spacing of $\Delta x = 0.04$ and $l_{\rm max}=4$, find a common AH in
the Brill-Lindquist data at $d=1.5$, but not at $d=1.6$ even after
doubling $l_{\rm max}$ and halving $\Delta x$.  Again, the
axisymmetric finder can determine the critical value to much higher
accuracy.  Running with the same resolution as that used for the
Misner data it finds a critical value of $d_*=1.532$ (consistent with
the value of Bishop {\em et al.}).  In Fig.~\ref{fig:brilin1} we
show our standard visual test for the position of the horizon for
$d=1.5$ (as found with the minimization algorithm with $l_{\rm
max}=8$).  For this horizon we find an area of $A \sim 2.0\times 10^2$.

\subsubsection{3 Black Hole data}

Both the Brill-Lindquist and Misner data generalize to an arbitrary
number of black holes with arbitrary masses.  This allows us to test
the apparent horizon finders on data that is not axisymmetric but is
still time-symmetric. The line element $dl^2$ is given by
(\ref{confla}) as before, and the Brill-Lindquist conformal factor for
$N$ black holes is
\begin{equation}
\psi=1 + \sum_{a=1}^N {m_a\over 2|\vec r - \vec r_a|}.
\label{brlipsi}
\end{equation}
The Misner conformal factor is obtained by adding an infinite sum
of ``mirror charges''~\cite{Anninos89}.

\begin{figure}
\epsfysize=3.4in \epsfbox{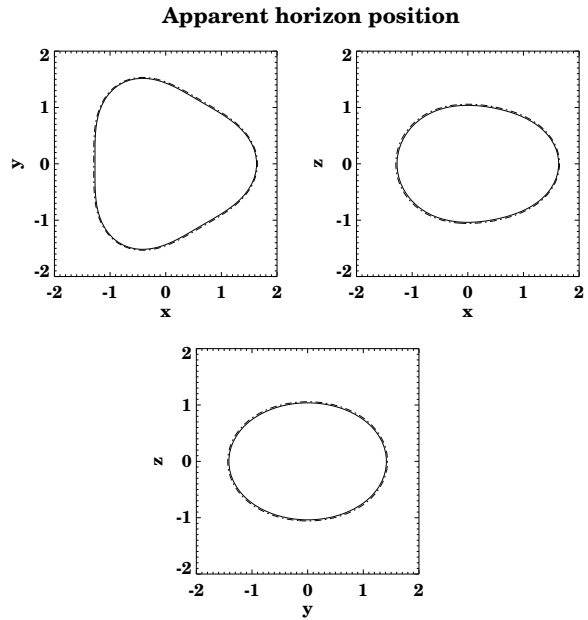}
\caption{Comparison of the AHs found with our two 3D finders
  for three equal mass Misner black-holes with $\mu=1.9$.  The solid
  line corresponds to the fast flow finder and the dotted line to the
  minimization finder.}
\label{fig:3misner1}
\end{figure}

\begin{figure}
\epsfysize=3in \epsfbox{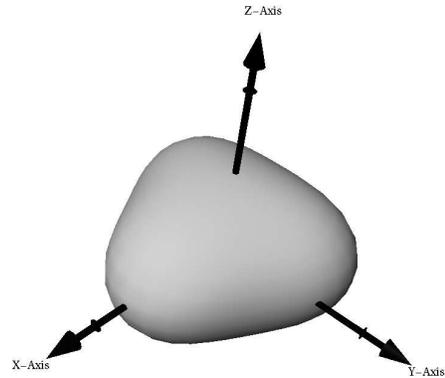}
\caption{3D representation of the coordinate location of the AH for
  three equal mass Misner black-holes with $\mu=1.9$}
\label{fig:3misner2}
\end{figure}

Nakamura {\em et al.}~\cite{Nakamura84} have tested a
three-dimensional AHF on a constellation of three equal mass black
holes of the Misner type -- two asymptotically flat regions joined by
three wormholes -- arranged in an equilateral triangle.  They
parameterize the family by $x=$ (coordinate side length of the
triangle) / (coordinate radius of the throats), and find a common
horizon for all three black holes for \mbox{$x \le 6.2$}.  We
parameterize the same family by $\mu$, which for this setup is related
to $x$ as \mbox{$\mu = \cosh^{-1}(x/\sqrt{3})$}, with $x=6.2$
corresponding to $\mu=1.9483$.  Both our 3D finders, working with a
resolution of $\Delta x = 0.05$ and using $l_{max}=6$, clearly find a
common horizon for $\mu=1.9$, and clearly do not find one for
$\mu=2.0$, so that we estimate \mbox{$\mu_* \in [1.9,2.0]$},
corresponding to \mbox{$x_* \in [5.9,6.5]$}.  Fig.~\ref{fig:3misner1}
shows a comparison of the AHs found with our two 3D finders for the
case $\mu=1.9$, using in both cases $l_{max}=6$ and a resolution of
$\Delta x = \Delta y = \Delta z = 0.1$.  Both finders find the same
surface to high accuracy.  Fig.~\ref{fig:3misner2} shows a 3D
representation of the coordinate location of this horizon.  The area
of this horizon was found to be $A\sim3.8\times10^2$.

\begin{figure}
\epsfysize=3.4in \epsfbox{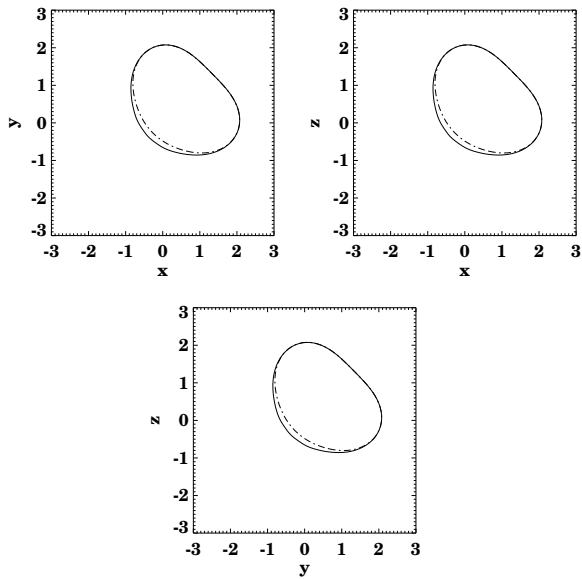}
\caption{Comparison of the AHs found with our two 3D finders
  for three equal mass Brill-Lindquist black-holes with $d=1.4$.  The
  solid line corresponds to the fast flow finder and the dotted line
  to the minimization finder.}
\label{fig:3brilin1}
\end{figure}

\begin{figure}
\epsfysize=3in \epsfbox{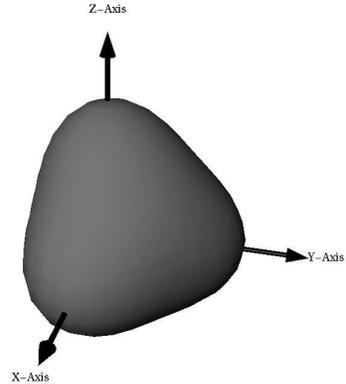}
\caption{3D representation of the coordinate location of the AH for
  three equal mass Brill-Lindquist black-holes with $d=1.4$}
\label{fig:3brilin2}
\end{figure}

Because Brill-Lindquist initial data are more easily obtained, we have
also looked for the maximal separation for Brill-Lindquist data for
the same setup. The three black holes are at coordinate locations
$(d,0,0)$, $(0,d,0)$ and $(0,0,d)$ in Cartesian coordinates, each with
a mass of one. This means that they are in an equilateral triangle of
side length $\sqrt{2}d$. The fast flow AHF, working in three
dimensions with a grid spacing of $\Delta x = \Delta y = \Delta z =
0.1$ and $l_{max}=12$, finds a common AH in the Brill-Lindquist data
at $d=1.4$, but not at $d=1.5$ even after doubling $l_{max}$ and
turning off the matrix inversion step described in~\cite{Gundlach97a}.
We therefore estimate \mbox{$d_* \in [1.4,1.5]$}, by criteria which
did give a correct bracketing for the 2 BH Misner and Brill-Lindquist
data, and the 3 BH Misner data.  The minimization algorithm, working
with the same resolution and with $l_{max}=6$, finds the same results
but is now painfully slow, taking about $25$ times longer than the
fast flow on exactly the same data.  This is because the black holes
have been placed in such a way that there are no reflection symmetries
on the coordinate planes, and this forces us to work with all
$\{l,m\}$ terms in the expansion.  In contrast, the 3 Misner black
holes above were set up on the $x-y$ plane and had reflection
symmetries both on the $x-y$ and $x-z$ planes, which allowed us to
eliminate many terms from the expansion, resulting in making the
minimization algorithm only $3$ times slower than the fast flow.  Of
course, we could have used the same configuration for the
Brill-Lindquist case, but we wanted to test our finders in a situation
with no special symmetries.  Fig.~\ref{fig:3brilin1} shows a
comparison of the AHs found with our two 3D finders for the case
$d=1.9$ using $l_{max}=6$.  Again, both finders find the same surface,
except close to the origin where there is a clear mismatch.  This
mismatch is a consequence of a lack of resolution in the spectral
decomposition, and disappears if we increase $l_{max}$ to $9$ (the
fast flow `horizon' moves in to lie on top of the original horizon
found with the minimization algorithm).  Fig.~\ref{fig:3brilin2} shows
a 3D representation of the coordinate location of the same horizon.
For this horizon we have found the area to be $A\sim4.5\times10^2$.

\subsection {Non time-symmetric black hole data}

All the data sets considered so far are time-symmetric, $K_{ij} = 0$.
In this case, Eq.~(\ref{eqn:horizon1}) defining marginally trapped
surfaces reduces to a minimal surface equation that involves only the
three-metric.  Allowing $K_{ij} \neq 0$ introduces a completely new
qualitative feature.  While minimal surface equations have been
studied a great deal in many different contexts, and while this is to
a lesser extent also true in the case of non-flat metrics, general
relativity with non-vanishing $K_{ij}$ introduces particular velocity
and potential terms about which little appears to be known.  For
example, see~\cite{Tod91} where it is pointed out that the natural
generalization of the mean curvature flow method does not define a
gradient flow and the area does not necessarily decrease.
Nevertheless, as stated in~\cite{Tod91}, there is good reason to hope
that such a flow algorithm will still converge.  Here we present
examples that demonstrate that all three finders are able to locate
apparent horizons in black hole space times with $K_{ij} \neq 0$.  Of
course, $K_{ij} \neq 0$ naturally arises during {\em evolution} of the
time-symmetric black hole data presented above, but we will not study
evolutions in this paper.

Analytically known black hole initial data which are not
time-symmetric include of course those obtained from slicing a single
Kerr black hole. These data are still axisymmetric, and the horizon is
a coordinate sphere. One can break the axisymmetry (of the data on a
slice, not the spacetime!) by boosting the Kerr black hole in a
direction not parallel to its angular momentum. This is done by
writing the Kerr metric in Kerr-Schild form
\begin{equation}
g_{\mu\nu} = \eta_{\mu\nu} + l_\mu l_\nu,
\end{equation}
where $\eta_{\mu\nu}$ is flat and $l_\mu$ is null (with respect to
both $\eta_{\mu\nu}$ and $g_{\mu\nu}$). In Cartesian coordinates
$(t,x,y,z)$, one applies a Lorentz transformation to $g_{\mu\nu}$, and
carries out a 3+1 split on the surface $t=0$. Explicit formulae are
given in
\cite{Matzner98a}. We have found the apparent horizon, and have
obtained good agreement between the minimization and fast flow
finders, for values for the dimensionless angular momentum of $0$,
$0.5$ and $0.8$, and for the same values of the boost speed, with all
combinations of these two parameters. Nevertheless, we felt that these
initial data still carried too much symmetry.

In the remainder of this section we will consider conformally flat
initial data for moving and spinning black holes that generalizes
Brill-Lindquist data~\cite{Brandt97b}. Similar tests could be carried
out for a generalization of Misner type data~\cite{Cook93}.  The
conformally flat line element is given by (\ref{confla}). The
extrinsic curvature is trace free, and therefore the momentum
constraint does not depend on the conformal factor.  An explicit
solution to the momentum constraint that characterize a single black
hole with given momentum $P^i$, and spin $S^i$ is
\begin{eqnarray}
   K^{ij}_{PS} &=& 
   \frac{3}{2r^2} (P^i n^j + P^j n^i - (g^{ij} - n^i n^j) P^k n_k)  
\nonumber \\
   && + \frac{3}{r^3} (\epsilon^{ikl} S_k n_l n^b 
                      +\epsilon^{jkl} S_k n_l n^i),
\label{KPJ}
\end{eqnarray}
where $n^i$ is the unit radial vector. Since the momentum constraint
is linear in $K_{ij}$, for $N$ black holes we can solve the momentum
constraint by
\begin{equation}
  K^{ij} = \sum_{a = 1}^{N} K_{PS(a)}^{ij},
\label{Ksol}
\end{equation}
where each term is defined by (\ref{KPJ}) with its own origin
$\vec{r}_{a}$, momentum $\vec{P}_{a}$, and spin $\vec{S}_{a}$.
These parameters correspond to the ADM quantities in the limit that
the separation of the holes is very large.

The Hamiltonian constraint is solved by splitting a regular function, $u$, 
from the conformal factor, compare (\ref{brlipsi}), 
\begin{eqnarray}
&&  \psi = u + 
\sum_{i=1}^{N} 
\frac{m_{a}}{2 \left|\vec{r}-\vec{r}_{a}\right|},
\label{brbrpsi} 
\\
&& \Delta_\delta u + \beta (1 + \alpha u)^{-7} = 0, 
\end{eqnarray}
where 
$\alpha = (\sum {m_{a}}/(2 \left|\vec{r}-\vec{r}_{a}\right|))^{-1}$, 
$\beta = \alpha^7 K^{ij}K_{ij}/8$, and $u\rightarrow1$ at infinity. 
A key feature of the ``puncture method'' is that
the resulting elliptic equation for $u$ can be implemented on $R^3$
despite the apparent singularities at the $\vec{r}_{a}$
\cite{Brandt97b}.


\subsubsection{One black hole data}

\begin{figure}
\epsfysize=8cm 
\centerline{\epsfbox{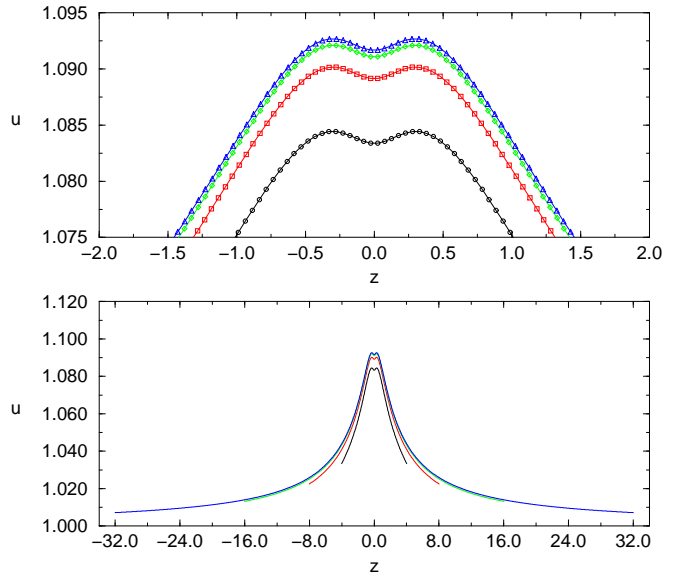}}
\caption{ 
  Regular part of the conformal factor, $u$, for a single black hole
  with $m = 1$ and $\vec P = (0,0,1)$ for various locations of the outer
  boundary. While the Robin boundary condition assumes a $1/r$
  fall-off, one clearly sees the influence of the $1/r^2$ terms.  
}
\label{fig:brbr0}
\end{figure}

\begin{figure}
\vspace{-3cm}
\epsfysize=8cm 
\centerline{\epsfbox{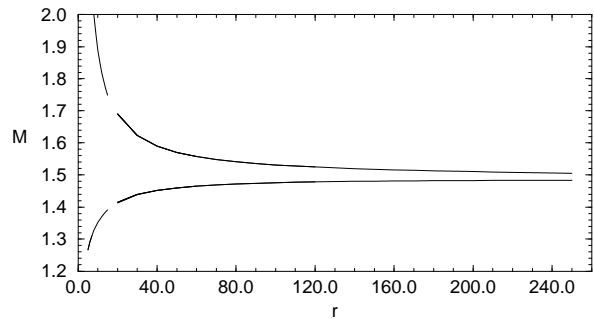}}
\caption{ 
  ADM mass integral (upper curve) and pseudo Schwarzschild mass (lower
  curve) for a single black hole with $m = 1$ and $\vec P = (0,0,1)$, data
  from the 3D solver with nested boxes and central resolution of $\Delta
  = 0.03125$. The leftmost piece of the curves is obtained for 5
  nested boxes of sizes $[-1,1]^3$ to $[-16,16]^3$, the other from 8
  nested boxes of sizes $[-2,2]^3$ to $[-256,256]^3$.
}
\label{fig:brbrm}
\end{figure}

First we consider a single black hole at the origin with mass $m=1$
and momentum aligned with the $z$-axis.  For non-vanishing net
momentum the fall-off as one approaches infinity is rather slow, and
one has to place the outer boundary of the grid sufficiently far away
to obtain, e.g., mass estimates close to the ADM mass (even though a
Robin boundary modeling a $1/r$ fall-off in the conformal factor was
used). In Fig.~\ref{fig:brbr0} we show the dependence of the regular
piece of the conformal factor, $u$, on the location of the outer
boundary for the 2D solver, $\Delta z=0.05$.  Note that $u = 1$ for
$\vec{P} = 0$.  Even though $\psi$ has a pole at $r = 0$, the location
of the outer boundary affects the location of the apparent horizon, which
in this case intersects the $z$-axis at 0.433 for the boundary at 4.0
and at 0.436 for the boundary at 32.0. 

In Fig.~\ref{fig:brbrm} we show how two different mass indicators
depend on radius for the 3D solver (we will discuss how we calculate
the masses in section~\ref{sec:brill} below). In 3D, some sort of
adaptivity is crucial for numerical efficiency. While the approach to
the mass at infinity is slow, perhaps surprisingly so considering that
the apparent horizon is at about $r = 0.5$, note that in general there
does not exist a concept of local mass that would allow one to compute
the mass at infinity at finite $r$.

\begin{figure}
  \epsfysize=3.4in \epsfbox{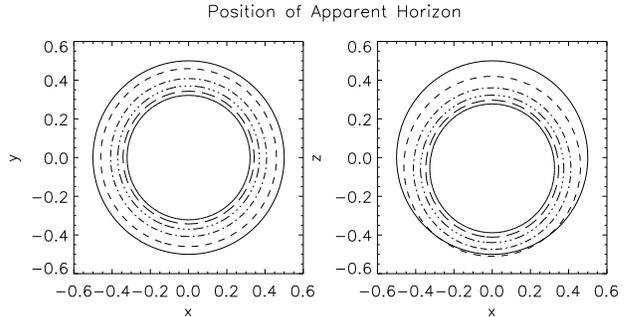} \vspace{-3.8cm}
\caption{ 
  Position of the apparent horizon for a single black hole at the
  origin with $m = 1$ and linear momentum in the $z$-direction.  For
  $P^z = 0.0$, the AH is a sphere of radius $0.5$.  For $P^z = 1.0$,
  $2.0$, $\ldots$, $5.0$, the AH shrinks and is shifted towards
  negative $z$ (in these coordinates).  In all cases we used $l_{max}
  = 8$ and the result for the two 3D finders cannot be distinguished.
  }
\label{fig:brbr1}
\end{figure}

In Fig.~\ref{fig:brbr1} we show results for the 3D apparent horizon
finders for a sequence of linear momenta aligned with the $z$-axis.
Here we put the outer boundary at $r = 16$ which appears reasonable
according to Fig.~\ref{fig:brbr0}. In order to achieve sufficient
resolution near the apparent horizon, the 3D data is computed on
five nested boxes with a refinement factor of 2, $[-16,16]^3$ to
$[-1,1]^3$, $64^3$ points each, with smallest grid spacing of $2/64 =
0.03125$.  To quantify the agreement between the two 3D finders, we
give various spectral coefficients for two resolutions in
Table~\ref{tab:brbr2}. 

Note that with increasing momentum the apparent horizon shrinks in
these coordinates. However, the surface area computed from the metric
increases: for $P^z = 0$, 1, 2, 3, 4, and 5, the surface areas
divided by $16\pi$ as found by the minimization algorithm at $\Delta =
0.0625$ are 1.000102, 1.19, 1.52, 1.85, 2.16, and 2.46, respectively.
 
Furthermore, the apparent horizon is offset from the origin in such a
way that it trails the ``motion'' of the center (see
also~\cite{Bruegmann97}).  This, of course, is a coordinate dependent
notion. For the event horizon, as opposed to the apparent horizon, one
would expect that it is displaced in the direction of the momentum
because an observer would find it harder to avoid falling into a black
hole that is moving towards her.

The trailing of the apparent horizon seems plausible by the following
argument. Note that the extrinsic curvature is odd under reversal of
momentum, $K_{ij}(-P) = - K_{ij}(P)$, while the conformal factor is
even, $\psi(-P) = \psi(P)$, since the extrinsic curvature enters into
the Hamiltonian constraint as $K_{ij}K^{ij}$. The expansion formula
(\ref{eqn:horizon1}) contains therefore an even term, $\nabla_i s^i$,
where the change in $\psi$ amounts to a symmetric deformation. The
remaining term $K_{ij}s^is^j$ (since $trK = 0$), is odd in $P^i$ {\em
  and} $n^i$. Since for $P^i=0$ and for concentric spheres of radius
$r$, the expansion $H(r)$ has a zero at $r=1/2$ with positive slope,
we expect to see the location of the horizon on the positive $z$-axis
to move to smaller $z$ with increasing $P^z > 0$. For a rigorous
argument one would have to take the non-locality of the minimal
surface equation into account.

The 2D results agree with the 3D results to within less than $1\%$,
which would not be visible in Fig.~\ref{fig:brbr1}. In this case the
initial data is obtained from an independent numerical code in 2D and 3D.
A simple test case which is independent of numerical error in the
initial data can be obtained by setting $u = 1$ for non-vanishing
$K_{ij}$. For the 2D data one can read off that for $u=1$ the
apparent horizon is almost exactly an ellipse with radius $0.495$ in
the $y$-direction, $0.499$ in the $z$-direction, and an offset in the
$z$-direction of $-0.061$.

\begin{table}
\begin{tabular}{c||r|r||r|r}
$m=0$
&
\multicolumn{2}{c||}{$64^3$, $\Delta = 0.03125$} 
&
\multicolumn{2}{c}{$32^3$, $\Delta = 0.06250$} 
\\ \hline
$l$ & fast flow & minim.\ & fast flow & minim.\   
\\ \hline
0 &  4.616$\times10^{-1}$ &  4.616$\times10^{-1}$ &  4.622$\times10^{-1}$ &  4.623$\times10^{-1}$ \\
1 & -2.601$\times10^{-2}$ & -2.602$\times10^{-2}$ & -2.624$\times10^{-2}$ & -2.615$\times10^{-2}$ \\
2 &  1.561$\times10^{-3}$ &  1.560$\times10^{-3}$ &  1.562$\times10^{-3}$ &  1.574$\times10^{-3}$ \\
3 & -9.020$\times10^{-5}$ & -8.749$\times10^{-5}$ & -1.251$\times10^{-4}$ & -8.437$\times10^{-5}$ \\
4 &  6.396$\times10^{-7}$ &  1.571$\times10^{-5}$ &  4.636$\times10^{-5}$ &  4.991$\times10^{-5}$
\end{tabular}
\caption{Spectral coefficients defining the apparent horizon for a
  single black hole with $m = 1$ and $\vec{P}=(0,0,1)$ for
  $l_{max} = 8$.}
\label{tab:brbr2}
\end{table}


\subsubsection{Two black hole data}

\begin{figure}
\epsfysize=3.4in \epsfbox{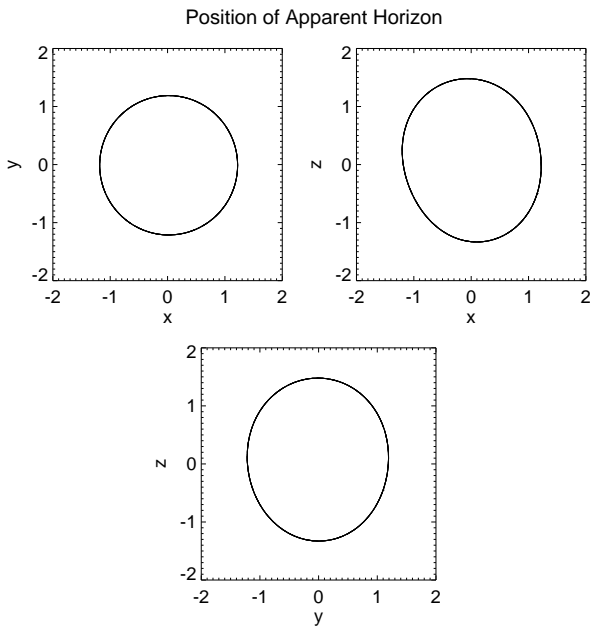}
\caption{Position of the apparent horizon for a particular binary black
  hole initial data set that is neither time-symmetric nor
  axisymmetric.  Results for the 3D finders for central resolution
  $\Delta = 0.1$ and $0.05$ and for $l_{max}=6$ and $8$ are plotted on
  top of each other.  The difference between the finders hardly shows.
  The momenta $\vec{P}_a$ are chosen such that in the $x-z$ plane the
  upper hole ``moves'' to the right, the lower to the left.}
\label{fig:brbr3}
\end{figure}

We consider one particular example for non-time-symmetric and
non-axisymmetric black hole binary initial data.  Such data was for
the first time evolved through a brief merger phase as indicated by
the location of the apparent horizon in \cite{Bruegmann97}.  Here we
compare the 3D finders for the example of \cite{Bruegmann97} but with
a separation such that there is one common outermost marginally
trapped surface already at $t=0$, Fig. \ref{fig:brbr3}: $m_1=1.5$,
$m_2=1$, $\vec{c}_{1,2}=(0,0,\pm 0.5)$, $\vec{P}_{1,2}=(\pm 2,0,0)$,
$\vec{S}_1=(-0.5,0.5,0)$, $\vec{S}_2=(0,1,1)$.  There are three grids
$[-12.8,12.8]^3$ to $[-3.2,3.2]$ with either $64^3$ or $128^3$ points
each, so that the central resolution is $\Delta = 0.1$ or $0.05$,
respectively.  Clearly, there is a very large parameter space to
study, and even more interestingly, one can also study how various
black hole data sets evolve through a merger, an issue that we hope to
address in a future publication.

\section{Brill wave spacetimes}
\label{sec:brill}

We finally turn to a rather difficult problem: determining whether or
not a horizon exists, and if so, determining its location, in a
numerically generated initial data set.  This will be a common
situation in simulations involving gravitational collapse of matter or
gravitational waves.  For this test, we turn to a sequence of pure
wave spacetimes, of a family originally considered by
Brill~\cite{Brill59}, that must be obtained numerically through
solutions to the constraint equations.  If the waves are strong
enough, an apparent horizon must be present~\cite{Beig91}, but at what
point it appears as one increases the wave strength, or where it will
be located, is unknown {\em a priori}.


\subsection{Initial data}

Brill's construction starts by considering an axisymmetric
metric of the form:
\begin{equation}
ds^2 = \Psi^4 \left[ e^{2q} \left( d\rho^2 + dz^2 \right)
+ \rho^2 d\phi^2 \right] .
\label{eqn:brillmetric}
\end{equation}

Where both $q$ and $\Psi$ are functions of $\{\rho,z\}$.  In order
to solve for $\Psi$, we first impose the condition of time symmetry,
which implies that the momentum constraints are identically satisfied.
We then chose a function $q$ and solve the Hamiltonian constraint,
which for the metric~(\ref{eqn:brillmetric}) becomes
\begin{equation}
\Delta_{\delta} \Psi + \frac{1}{4} \, (q_{,\rho\rho}
+ q_{,zz})\, \Psi = 0 ,
\label{eqn:brillham}
\end{equation}
where $\Delta_{\delta}$ is the flat space Laplacian.

The function $q$ is almost completely arbitrary, apart from the fact
that it must satisfy the following boundary conditions
\begin{eqnarray}
q \left|_{\rho=0} \right. &=& 0 ,\\
\partial^n_\rho q \left|_{\rho=0} \right. &=& 0 \qquad \mbox{for odd $n$},\\
q \left|_{r\rightarrow\infty} \right. &=& O \left( r^{-2} \right) .
\end{eqnarray}

Once a function $q$ has been chosen, all that is left for one to do is
solve the elliptic equation~(\ref{eqn:brillham}) numerically.  This
can be done in a variety of ways.  Here we use two independent
elliptic solvers: an axisymmetric solver based on a semi-coarsening
multi-grid solver, and a multi-grid fully 3D solver (Bernd
Br\"{u}gmann's BAM) hooked up to the Cactus~\cite{Bona98a} code
recently developed at the Albert-Einstein-Institut.  Having two
independent solvers with different methods, in different coordinate
systems, and in different dimensions has allowed us to cross check our
results and has increased our confidence in our solution of the
elliptic problem.

Different forms of the function $q$ have been used by different
authors~\cite{Eppley77,Eppley79,Holz93,Shibata97b}.  Here we will
consider two such forms, the first one introduced by Eppley in his
pioneering work on Brill waves in the
seventies~\cite{Eppley77,Eppley79}, and the second one introduced by
Holz {\em et al.} in the early nineties~\cite{Holz93}.

Before moving on to horizon finding in pure wave spacetimes, we first
comment on how to calculate the gravitational mass of a given initial
data set.  Finding the gravitational mass is a very useful tool in
testing the accuracy of our initial data.  It provides us with a
single number that can be easily compared for different initial data
solvers and is a good indicator of the strength of a gravitational
wave.  For strong wave spacetimes that collapse to a black hole, the
difference between the initial mass of the wave and the mass of the
final black hole is a good indicator of the percentage of energy that
was radiated out to infinity.  We have several ways of calculating the
initial mass of our spacetimes.  The first method is to use the ADM
mass~\cite{Wald84}
\begin{equation}
M = \frac{1}{16 \pi} \lim_{r\rightarrow\infty} \oint g^{ij} g^{mn}
\left( g_{in,j} - g_{ij,n} \right) \sqrt{g} \; dS_m 
\label{eq:ADMmass1}
\end{equation}
in appropriate coordinates.

Of course, our numerical grid does not extend all the way to infinity,
so in practice we evaluate the integral at a series of different
finite radii and look at its behavior as $r$ increases.  As it turns
out, the mass calculated using Eq.~(\ref{eq:ADMmass1}) converges only
very slowly with $r$ even for a simple Schwarzschild black hole.  A
better way of calculating the mass uses the fact that for large $r$
(but still small enough to be inside the computational grid) the
function $q$ becomes essentially zero and the metric is conformally
flat.  For conformally flat metrics the ADM mass can be rewritten
as~\cite{Omurchadha74}
\begin{equation}
M = - \frac{1}{2 \pi} \lim_{r\rightarrow\infty} \oint \vec{\nabla}
 \Psi \cdot d \vec{S} .
\label{eq:ADMmass2}
\end{equation}

Equations~(\ref{eq:ADMmass1}) and~(\ref{eq:ADMmass2}) are only
equivalent in the limit of infinite radius, but it turns out that for
a Schwarzschild black hole in isotropic coordinates,
Eq.~(\ref{eq:ADMmass2}) gives in fact the correct mass at {\em any
  finite radius}.  Since once we are in the region where $q$ is very
small the metric of our Brill wave solutions approaches the
Schwarzschild metric rapidly, one finds that the masses obtained by
using~(\ref{eq:ADMmass2}) converge very fast as $r$ increases.

A final way of calculating the mass is what we call the `pseudo
Schwarzschild mass'.  This mass estimate is obtained by first
finding the areal (Schwarzschild) radius $R$ of a series of
coordinate spheres, finding the correspondent metric component $g_{RR}$
(averaged over the coordinate sphere) and then defining:
\begin{equation}
M = \frac{R}{2} \left( 1 - \frac{1}{g_{RR}} \right) .
\label{eq:ADMmass3}
\end{equation}

In practice we find that for the spacetimes studied below, the mass
indicator~(\ref{eq:ADMmass3}) also converges very rapidly with $r$.


\subsection{Eppley data}

Eppley considered a function $q$ of the general form~\cite{Eppley77}
\begin{equation}
q = a \; \frac{\rho^2}{1 + (r/\lambda)^n},
\label{eqn:eppleyq}
\end{equation}
where $a,\lambda$ are constants, $r^2 = \rho^2 + z^2$ and $n \geq 4$.
Notice here that, for odd $n$ the function $q$ is not
completely smooth at the origin.  Nevertheless, Eppley considered
mainly the particular case $\lambda=1$, $n=5$, and in order to compare
with his results we will do the same here.

Before looking for horizons, we must first convince ourselves that we
can solve for the conformal factor $\Psi$ correctly, that is, that we
can construct good initial data.  Our approach here is to solve the
initial value problem independently in axisymmetry and in full 3D and
compare both results with those of Eppley.  In particular, we will
look at the extracted masses for a sequence of solutions with
increasing amplitudes $a$.  For our axisymmetric initial data we have
used a grid of $800^2$ points with a resolution of $\Delta \rho =
\Delta z = 0.03125$, and for the 3D data a grid of $131^3$ points with
a resolution of $\Delta x = \Delta y = \Delta z =0.08$.  In
Table~\ref{tab:epp_mass} we tabulate the values of the masses that we
find for different wave amplitudes.  The masses that we report here as
those obtained by Eppley were read off (by us) from Fig.~1 in
Ref.~\cite{Eppley77} (modulo a conventional factor of 2 in the
amplitude $a$) and are thus not very accurate.  The error estimates
for the 2D calculations were determined from the difference of the
masses obtained by looking at the falloff of the conformal factor
along the $z$ and $\rho$ axis.

\begin{figure} \def\epsfsize#1#2{0.5#1}
\centerline{\epsfbox{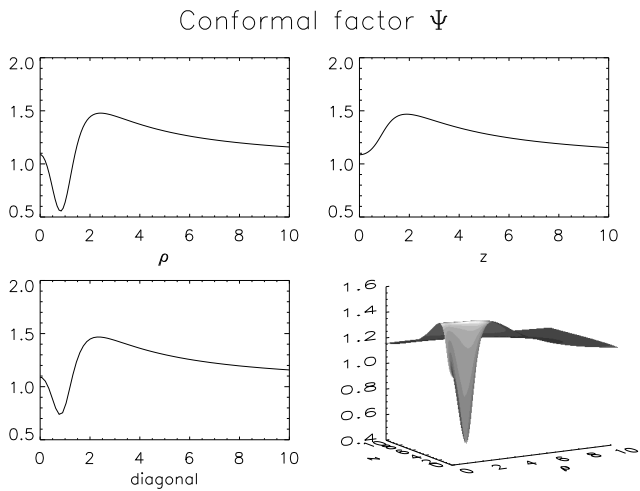}}
\caption{Conformal factor $\Psi(\rho,z)$ for Brill wave initial data
  with a source function $q$ of the Eppley type and an amplitude
  $a=10$.}
\label{fig:epp_psi}
\end{figure}

From Table~\ref{tab:epp_mass} we can see that the masses obtained with
our axisymmetric and 3D elliptic solvers agree remarkably well between
themselves, but are generally different from those reported by Eppley.
For low amplitudes, Eppley's masses are lower than those that we
find.  For an amplitude $a\simeq5$, Eppley's mass and ours coincide,
but for larger amplitudes Eppley's masses grow much faster.  In fact,
Eppley reports that for $a\simeq8$, the geometry pinches off (the
conformal factor has a zero) and the mass becomes infinite but we see
no evidence of such behavior.  Since our two {\em independent} initial
data solvers agree so well with each other, we are forced to conclude
that there must have been something wrong in Eppley's calculations.
It must be pointed out here that Eppley makes a very strong point of
trying to calculate the masses correctly, so we must conclude that the
error must have been in his solution for the conformal factor.
As an example of our initial data, in Fig.~\ref{fig:epp_psi} we show
the conformal factor $\Psi(\rho,z)$ for the case $a=10$.

\begin{table}
\begin{tabular}{c|c|c|c}
$a$ & M (2D) & M (3D) & M (Eppley) \\ \hline
1   & $(4.8\pm 0.1)\times 10^{-2}$ & $5.0\times10^{-2}$ & $(3.6\pm 0.2)\times10^{-2}$ \\
2   & $(1.74\pm 0.02)\times 10^{-1}$ & $1.8\times10^{-1}$ & $(1.1\pm.1)\times10^{-1}$\\
5   & $(8.83\pm 0.07)\times 10^{-1}$ & $8.9\times10^{-1}$ & $(9\pm 0.5)\times10^{-1}$ \\
10  & $3.22\pm 0.02$                 & 3.2                & - \\
12  & $4.85\pm 0.02$                 & 4.9                & -
\end{tabular}
\caption{Masses for Brill wave initial data with a source function
$q$ of the Eppley~{\em et al.} type.  Notice how our 2D and 3D solvers
agree remarkably well between each other, but disagree with Eppley's
results.}
\label{tab:epp_mass}
\end{table}

Having constructed the initial data, we will now look for the presence
of AHs.  We have studied a series of solutions with increasing
amplitudes with all three of our AHFs.  Both our 3D finders find that
an AH first appears for a critical amplitude $a_*\in[10.8,10.9]$.  For
amplitudes above $a_*$ we can find two horizons using the minimization
algorithm (the fast flow algorithm is not designed to look for inner
horizons).  The 2D AHF can pin-point the value of $a*$ more precisely
to $a_*\in[10.86,10.87]$.  As we can see, the agreement in the
value of $a_*$ is remarkable.

\begin{figure}
\epsfysize=2.2in \epsfbox{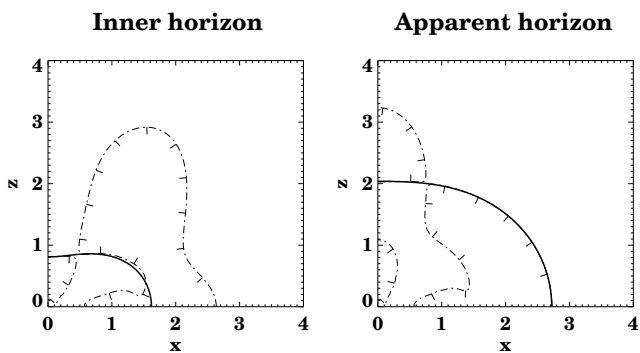}
\caption{Position of the inner horizon and AH on the $x-z$ plane
  for a Brill wave initial data set using Eppley's $q$ function
  with $a=12$.  The solid lines are the positions of the horizons,
  and the dotted lines are the zeroes of the expansion $H$.}
\label{fig:eppley1}
\end{figure}

\begin{figure}
\epsfysize=3in \epsfbox{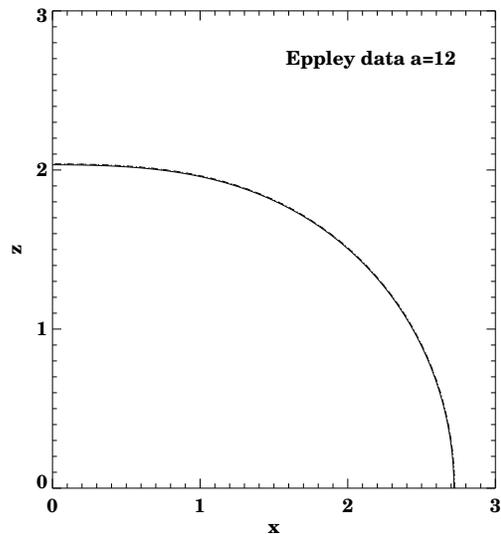}
\caption{Comparison of the AHs found with our three finders for a
  Brill wave initial data set using Eppley's $q$ function with $a=12$.
  The solid line corresponds to the 2D finder, the dotted line to the
  fast flow finder, and the dashed line to the minimization finder.
  All three lines lie on top of each other.}
\label{fig:eppley2}
\end{figure}

It is of course not enough to agree on the value of the critical
amplitude above which AHs appear.  We also need to compare the
positions of the horizons found by the different AHFs. We have done
this for many different amplitudes and found good agreement between
our three AHFs. In Fig.~\ref{fig:eppley1} we show our standard visual
test for the position of both the inner horizon and the AH for the
particular case of $a=12$.  Notice how in both cases the horizons
coincide with zeroes of the expansion.  Fig.~\ref{fig:eppley2} shows a
comparison of the position of the AH found with the different
algorithms.  Again, all three finders locate the same surface.  In
this case we find that the area of the AH is $A\sim1.1\times 10^3$.
It is interesting to see whether this result agrees with the Penrose
inequality~\cite{Penrose73} $16 \pi M^2/A \geq 1$.  From
Table~\ref{tab:epp_mass} we see that the ADM mass in this case in $M
\sim 4.85$, which implies $16 \pi M^2/A \sim 1.07$, so the inequality
is comfortably satisfied.

\subsection{Holz data}

Holz {\em et al.} considered a Brill wave source function $q$
of the form
\begin{equation}
q = a \; \rho^2 \; e^{-r^2} .
\label{eqn:holzq}
\end{equation}
This form of the function $q$ is perfectly regular at the origin.  An
almost identical form of the function $q$ was also recently considered
by Shibata~\cite{Shibata97b}
\begin{equation}
q_{\rm Sh} = {a\over 2} \; \rho^2 \; e^{-r^2/2} .
\label{eqn:shibataq}
\end{equation}
(Note that Shibata has a different convention for $q$; this is our
convention.) One can see that Shibata's results can in fact be
compared directly with those of Holz {\em et al.} because the two
metrics differ only by a factor of $\sqrt{2}$ in the coordinates (and
therefore also in the ADM mass), which on rescaling can be
absorbed into the conformal factor (notice that multiplying the
conformal factor by a constant does not affect the equation for a
horizon~(\ref{eqn:horizon1}) for vanishing extrinsic curvature).  The
strength parameter $a$ is the same in both cases.

\begin{figure} \def\epsfsize#1#2{0.5#1}
\centerline{\epsfbox{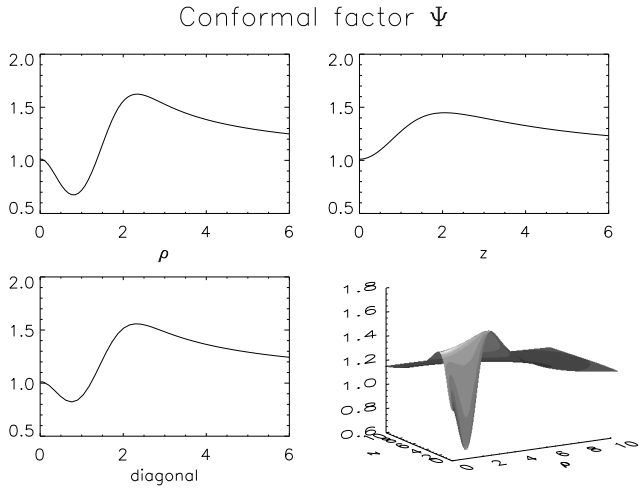}}
\caption{Conformal factor $\Psi(\rho,z)$ for Brill wave initial data
  with a source function $q$ of the Holz~{\em et al.} type and an
  amplitude $a=10$.}
\label{fig:holz_psi}
\end{figure}

Again, before looking for horizons, we will first test our initial
data solvers by comparing the solutions of our 2D and 3D elliptic
solvers with the results of Holz {\em et al.}~\cite{Holz93,Holz98x}.
As before, the 2D initial data was calculated using $800\times800$
grid points and a resolution of $\Delta z = \Delta \rho = 0.03125$ and
the 3D data using $131^3$ grid points with a resolution of $\Delta x =
\Delta y = \Delta z = 0.06$.  The values of the masses we find for
different wave amplitudes can be seen in Table~\ref{tab:holz_mass}.
Notice how the values extracted from the 2D and 3D data agree
remarkably well both between themselves and with the masses of Holz
{\em et al.}  This gives us great confidence in the accuracy of our
initial data.  Fig.~\ref{fig:holz_psi} shows in particular the
conformal factor $\Psi(\rho,z)$ for the case $a=10$.

\begin{table}
\begin{tabular}{c|c|c|c}
$a$ & M (2D) & M (3D) & M (Holz {\em et al.}) \\ \hline
1   & $(3.38 \pm .04)\times10^{-2}$   & $3.4\times10^{-2}$ & $3.40\times10^{-2}$ \\
2   & $(1.262 \pm .009)\times10^{-1}$ & $1.3\times10^{-1}$ & $1.27\times10^{-1}$ \\
5   & $(6.96 \pm .03)\times10^{-1}$   & $7.0\times10^{-1}$ & $7.00\times10^{-1}$ \\
10  & $2.912 \pm .008$                & 2.9                & 2.91 \\
12  & $4.67 \pm 0.01$                 & 4.7                & 4.68
\end{tabular}
\caption{Masses for Brill wave initial data with a source function
$q$ of the Holz~{\em et al.} type.}
\label{tab:holz_mass}
\end{table}

Although we agree with Holz {\em et al.} on the initial data, we
disagree on the AHs. Our three AHF agree among themselves, but
disagree with the results reported in Ref.~\cite{Holz93}.  Holz~{\em
et al.} claim that an AH first appears for a critical amplitude
$a_*=7.5$, and for larger amplitudes they can find two horizons.  Our
own results are qualitatively similar: a horizon first appears for a
given critical amplitude, and above that we can always find two
horizons.  However, the value for that critical amplitude is
different.  Both our 3D finders indicate that $a_*\in[11.8,11.85]$,
while the our 2D finder limits the interval to $a_*\in[11.81,11.82]$.
Shibata, on the other hand, finds that the first AH appears for
$a\sim12$, in complete agreement with our results.  The mass of the
solution corresponding to the critical amplitude turns out to be
$M\simeq4.5$.

\begin{figure}
\epsfysize=2.2in \epsfbox{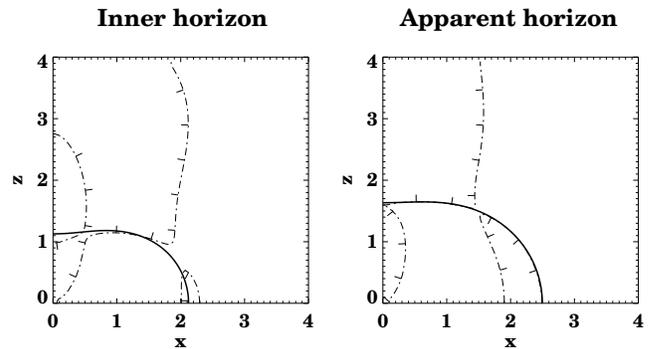}
\caption{Position of the inner horizon and AH on the $x-z$ plane
  for a Brill wave initial data set using Holz' $q$ function
  with $a=12$.  The solid lines are the positions of the horizons,
  and the dotted lines are the zeroes of the expansion $H$.}
\label{fig:holz1}
\end{figure}

\begin{figure}
\epsfysize=3in \epsfbox{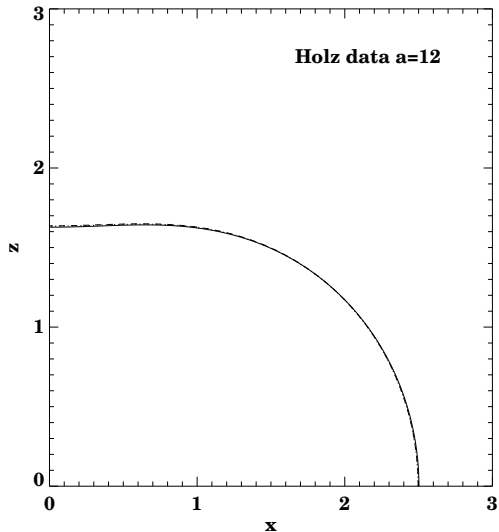}
\caption{Comparison of the AHs found with our three finders for a
  Brill wave initial data set using Holz' $q$ function with $a=12$.
  The solid line corresponds to the 2D finder, the dotted line to the
  fast flow finder, and the dashed line to the minimization finder.
  All three lines lie on top of each other.}
\label{fig:holz2}
\end{figure}

In Fig.~\ref{fig:holz1} we show again our standard visual test for the
position the inner and apparent horizons for the particular case of
$a=12$. Fig.~\ref{fig:holz2} shows a comparison of the position of the
AH found with our three different finders.  The area of the horizon in
this case turns out to be \mbox{$A \sim 1.1 \times 10^3$}.  From
Table~\ref{tab:holz_mass} we see that in this case the ADM mass is $M
\sim 4.67$, from which we find that $16 \pi M^2/A \sim 0.997$.  The
Penrose inequality appears to be slightly violated, but this small
violation could easily be caused by the inaccuracies in the
determination of both the mass and the area.

\section{Conclusions}

In this paper we have developed a large number of test-beds for 
apparent horizon finding algorithms in numerical relativity, and have 
applied them to three algorithms of ours.  There were several goals of 
this study.  First, in preparation for studies of a wide variety of 
datasets it was important to verify that our algorithms are robust and 
that their coding is correct.  As a result of this extensive testing 
we are now very confident that our algorithms, which had to be 
refined in several cases to find difficult but known horizons, are 
correct and robust.  Second, through this study we have developed an 
extensive series of quantitative tests for both developing and 
validating future algorithms, that should be useful to the community.  
Third, we used these refined horizon finding algorithms to study a 
series of interesting black hole and gravitational wave initial data 
sets in preparation for numerical evolutions.

For the purpose of validating our own algorithms, we have repeated
virtually all {\it quantitative} test-beds in the literature that have
come to our attention. Here it was important to have some simple
numbers that can be compared directly. For two of these test-beds
(Brill wave data), our results disagree with the published results. We
are confident that our results are the correct ones both because our
three finders agree with each other, and because they agree with the
literature for the other published test-beds.

At our disposal we had three apparent horizon finders, two independent 
ones in three space dimensions without symmetries, and one limited to 
axisymmetry.  Both 3D finders are an integral part of the Cactus 
numerical relativity framework~\cite{Bona98b}, so that all the 
test-bed data can be calculated, examined for horizons with either of 
the two finders, and evolved forward in time within the same code, 
just by changing parameters.  The 2D finder was also linked up with a 
2D initial data solver.  While the fast flow 3D AH finder is generally 
much faster than the minimization routine, it was crucial for the 
validation process to have all three available to work on the same 
data.  We stress that Powell's method, which was used in the 
minimization algorithm, is probably not the best for the minimization, 
but it is common and readily available for testing the basic AH 
finding strategy; more sophisticated minimization algorithms could 
accelerate the minimization-based AH finder.  The speed of the finders 
will be crucial in determining whether full scale numerical evolutions 
can be practically carried out or not, and even the present generation 
finders can be taxing in terms of computational time.  The 2D initial 
data solver and apparent horizon finder, although limited to 
axisymmetric data, had the advantage of allowing for much higher 
numerical precision, thus giving us more confidence yet in our results.

We want to emphasize the importance of proper validation 
through test-beds, of any new algorithm, especially for analysis tools 
such as horizon finders for which standard techniques such as 
convergence tests will generally not reveal algorithm deficiencies or 
coding errors.  Validation on simple examples with lower symmetry, or 
with analytically known horizons, is important but simply not 
sufficient.  One could almost establish a variant of Murphy's law 
stating that every new test-bed calculation that is in some way more 
generic than previous ones will reveal a new deficiency in a given 
finder.  As AHs are not known for completely general data sets, being 
able to test two or three totally independent finders on the same 
initial data was crucial to our development process, right down to the 
process of eliminating typos in this paper.  On the one hand, we have 
included some physically interesting data sets (black holes and Brill 
waves), on the other hand we would like to stress that for the sole 
purpose of testing an AHF, one can run it on data that do not obey the 
constraints (``bowl'' spacetimes).

Finding AHs in data without symmetries remains a very difficult
problem.  For generic data, which are not time-symmetric, no algorithm
is known to always work. Essentially, the problem is highly nonlinear,
and can be made arbitrarily difficult with sufficiently ``bad'' data
(a typical example of such behavior was discussed for the ``bowl''
spacetimes).  In fact there may not be any ``best'' algorithm that is
both fast and robust at the same time. Further, in different cases one
algorithm may work better than another, or vice versa.  For these
reasons we continue to use our various AH finders to confirm results.

As an aid to future development and validation efforts, in this paper
we have given simple numbers (critical separations, horizon areas and
ADM masses) for families of initial data that should provide useful
and quantitative testbed for other groups.  However, in our opinion,
validation must also include detailed comparison of the entire shape
of the candidate AH with other algorithms, for data which have no
symmetries at all.

The horizon finders developed and refined on these spacetimes are
presently being applied to evolutions of some of these datasets.
These results will be presented in future publications.


\acknowledgements We are indebted to many colleagues for numerous
discussions and email exchanges during the course of this work.  In
particular, Gabrielle~Allen, Daniel~Holz, Sascha~Husa,
Niall~O'Murchadha and Wai-Mo~Suen provided valuable input and insights
into the results at preliminary stages of this work. We also want to
thank Werner~Benger for helping us with the visualization of our 3D
data.  This work was supported by the Albert-Einstein-Institut, by 
NCSA, and by UIB.


\bibliographystyle{prsty}
\bibliography{bibtex/references}

\begin{thebibliography}{10}

\bibitem{Anninos94f}
P. Anninos {\it et~al.}, Phys. Rev. Lett. {\bf 74},  630  (1995).

\bibitem{Libson94a}
J. Libson {\it et~al.}, Phys. Rev. D {\bf 53},  4335  (1996).

\bibitem{Hughes94a}
S. Hughes {\it et~al.}, Phys. Rev. D {\bf 49},  4004  (1994).

\bibitem{Matzner95a}
R. Matzner {\it et~al.}, Science {\bf 270},  941  (1995).

\bibitem{Masso95a}
J. Mass\'o, E. Seidel, W.-M. Suen, and P. Walker, gr-qc/9804059. Submitted to
  Phys. Rev. D  (1998).

\bibitem{Shapiro95a}
S. Shapiro, S. Teukolsky, and J. Winicour, Phys. Rev. D {\bf 52},    (1995).

\bibitem{Hawking73a}
S.~W. Hawking and G.~F.~R. Ellis, {\em The Large Scale Structure of Spacetime}
  (Cambridge University Press, Cambridge, England, 1973).

\bibitem{Seidel92a}
E. Seidel and W.-M. Suen, Phys. Rev. Lett. {\bf 69},  1845  (1992).

\bibitem{Cook97a}
G.~B. Cook {\it et~al.}, Phys. Rev. Lett {\bf 80},  2512  (1998).

\bibitem{Gundlach98a}
C. Gundlach and P. Walker,   (1998), in preparation.

\bibitem{Anninos93a}
P. Anninos {\it et~al.}, Phys. Rev. D {\bf 50},  3801  (1994).

\bibitem{Anninos95c}
P. Anninos {\it et~al.}, Australian Journal of Physics {\bf 48},  1027  (1995).

\bibitem{Anninos93d}
P. Anninos {\it et~al.}, IEEE Computer Graphics and Applications {\bf 13},  12
  (1993).

\bibitem{Cadez74}
A. \v{C}ade\v{z}, Ann. Phys. {\bf 83},  449  (1974).

\bibitem{Eppley77}
K. Eppley, Phys. Rev. D {\bf 16},  1609  (1977).

\bibitem{Bishop82}
N.~T. Bishop, Gen. Rel. Grav. {\bf 14},  717  (1982).

\bibitem{Nakamura84}
T. Nakamura, Y. Kojima, and K. Oohara, Phys. Lett. {\bf 106A},  235  (1984).

\bibitem{Cook90a}
G. Cook and J.~W. York, Phys. Rev. D {\bf 41},  1077  (1990).

\bibitem{Kemball91a}
A.~J. Kemball and N.~T. Bishop, Class. Quantum Grav. {\bf 8},  1361  (1991).

\bibitem{Libson94b}
P. Anninos {\it et~al.}, Phys. Rev. D {\bf 58},  024003  (1998).

\bibitem{Libson95a}
J. Libson, J. Mass\'o, E. Seidel, and W.-M. Suen,  in {\em The Seventh Marcel
  Grossmann Meeting: On Recent Developments in Theoretical and Experimental
  General Relativity, Gravitation, and Relativistic Field Theories}, edited by
  R.~T. Jantzen, G.~M. Keiser, and R. Ruffini (World Scientific, Singapore,
  1996), p.\ 631.

\bibitem{Thornburg95}
J. Thornburg, Phys. Rev. D {\bf 54},  4899  (1996).

\bibitem{Baumgarte96}
T.~W. Baumgarte {\it et~al.}, Physical Review D {\bf 54},  4849  (1996).

\bibitem{Gundlach97a}
C. Gundlach, Phys. Rev. D {\bf 57},  863  (1998), gr-qc/9707050.

\bibitem{Shibata97a}
M. Shibata, Phys. Rev. D {\bf 55},  2002  (1997).

\bibitem{York89}
J. York,  in {\em Frontiers in Numerical Relativity}, edited by C. Evans, L.
  Finn, and D. Hobill (Cambridge University Press, Cambridge, England, 1989),
  pp.\ 89--109.

\bibitem{Brill63}
D.~S. Brill and R.~W. Lindquist, Phys. Rev. {\bf 131},  471  (1963).

\bibitem{Libson93a}
J. Libson,  in {\em Numerical Relativity Conference}, edited by P. Laguna
  (PUBLISHER, ADDRESS, 1993).

\bibitem{Camarda97a}
K. Camarda, Ph.D. thesis, University of Illinois at Urbana-Champaign, Urbana,
  Illinois, 1998.

\bibitem{Camarda97c}
K. Camarda and E. Seidel,   , gr-qc/9805099. Submitted to Physical Review D.

\bibitem{Press86}
W.~H. Press, B.~P. Flannery, S.~A. Teukolsky, and W.~T. Vetterling, {\em
  Numerical Recipes} (Cambridge University Press, Cambridge, England, 1986).

\bibitem{Bona98b}
C. Bona, J. Mass\'o, E. Seidel, and P. Walker,   (1998), gr-qc/9804065.
  Submitted to Physical Review D.

\bibitem{Tod91}
K.~P. Tod, Class. Quan. Grav. {\bf 8},  L115  (1991).

\bibitem{Anninos94c}
P. Anninos {\it et~al.}, Phys. Rev. D {\bf 52},  2059  (1995).

\bibitem{Misner60}
C. Misner, Phys. Rev. {\bf 118},  1110  (1960).

\bibitem{Hahn64}
S.~G. Hahn and R.~W. Lindquist, Ann. Phys. {\bf 29},  304  (1964).

\bibitem{Smarr76}
L. Smarr, A. \v{C}ade\v{z}, B. DeWitt, and K. Eppley, Phys. Rev. D {\bf 14},
  2443  (1976).

\bibitem{Anninos93b}
P. Anninos {\it et~al.}, Phys. Rev. Lett. {\bf 71},  2851  (1993).

\bibitem{Anninos94b}
P. Anninos {\it et~al.}, Phys. Rev. D {\bf 52},  2044  (1995).

\bibitem{Anninos89}
P. Anninos, Ph.D. thesis, Drexel University, 1989.

\bibitem{Matzner98a}
R.~A. Matzner, M.~F. Huq, and D. Shoemaker, to appear in Phys. Rev. D  (1998).

\bibitem{Brandt97b}
S. Brandt and B. Br\"ugmann, Phys. Rev. Lett. {\bf 78},  3606  (1997).

\bibitem{Cook93}
G.~B. Cook {\it et~al.}, Phys. Rev. D {\bf 47},  1471  (1993).

\bibitem{Bruegmann97}
B. Br\"ugmann,   (1997), gr-qc/9708035.

\bibitem{Brill59}
D.~S. Brill, Ann. Phys. {\bf 7},  466  (1959).

\bibitem{Beig91}
R. Beig and N. O'Murchadha, Phys. Rev. Lett. {\bf 66},  2421  (1991).

\bibitem{Bona98a}
C. Bona, J. Carot, and J. Mass\'o, In preparation  (1998).

\bibitem{Eppley79}
K. Eppley,  in {\em Sources of Gravitational Radiation}, edited by L. Smarr
  (Cambridge University Press, Cambridge, England, 1979), p.\ 275.

\bibitem{Holz93}
D.~E. Holz, W.~A. Miller, M. Wakano, and J.~A. Wheeler,  in {\em Directions in
  General Relativity: Proceedings of the 1993 International Symposium,
  Maryland; Papers in honor of Dieter Brill}, edited by B.~L. Hu and T.
  Jacobson (Cambridge University Press, Cambridge, England, 1993).

\bibitem{Shibata97b}
M. Shibata, Phys. Rev. D {\bf 55},  7529  (1997).

\bibitem{Wald84}
R.~M. Wald, {\em General Relativity} (The University of Chicago Press, Chicago,
  1984).

\bibitem{Omurchadha74}
N. O'Murchadha and J. York, Phys. Rev. D {\bf 10},  2345  (1974).

\bibitem{Penrose73}
R. Penrose, Ann. N.Y. Acad. Sci. {\bf 224},  125  (1973).

\bibitem{Holz98x}
D. Holz, private communication (unpublished).

\end{thebibliography}

\end{document}